\documentclass[prl, onecolumn, a4paper, superscriptaddress, showpacs, amsmath,
amssymb, nofootinbib, floatfix]{revtex4}

\usepackage[utf8]{inputenc}
\usepackage[english]{babel}
\usepackage{graphicx}
\usepackage{color}

\newcommand{\rmd}{\mathrm d}

\newcommand{\BM}[1]{\mbox{\boldmath$#1$}}
\newcommand{\Bm}[1]{\mbox{\scriptsize\boldmath$#1$}}
\newcommand{\Bl}{{\BM l}}
\newcommand{\Ba}{{\BM a}}
\newcommand{\Bb}{{\BM b}}
\newcommand{\Bu}{{\BM u}}
\newcommand{\Br}{{\BM r}}
\newcommand{\Bn}{{\BM n}}
\newcommand{\Bd}{{\BM d}}
\newcommand{\Be}{{\BM e}}
\newcommand{\Bg}{{\BM g}}
\newcommand{\Bk}{{\BM k}}
\newcommand{\BI}{{\BM I}}
\newcommand{\BC}{{\BM C}}
\newcommand{\BG}{{\BM G}}
\newcommand{\BT}{{\BM T}}
\newcommand{\BD}{{\BM D}}
\newcommand{\BR}{{\BM {\cal R}}}
\newcommand{\BE}{{\BM {\cal E}}}
\newcommand{\Bsigma}{{\BM \sigma}}
\newcommand{\Bkappa}{{\BM \kappa}}
\newcommand{\Bepsilon}{{\BM \epsilon}}
\newcommand{\Balpha}{{\BM \alpha}}
\newcommand{\Brho}{{\BM \rho}}
\newcommand{\BMM}{{\BM M}}

\newcommand \Michael[1]{\bgroup\noindent[\textcolor{blue}{\textbf{Michael}: #1}]\egroup\ignorespacesafterend}

\begin{document}

\title{Local density approximation for the energy functional of three-dimensional dislocation systems}

\author{M. Zaiser}%
\email[]{michael.zaiser@fau.de}
\affiliation{Institute for Materials Simulation WW8, FAU University of
Erlangen-Nuremberg, Germany}

\date{\today}


\begin{abstract}
The elastic energy functional of a system of discrete dislocation lines is well known from dislocation theory. 
In this paper we demonstrate how the discrete functional can be used to systematically derive approximations 
which express the elastic energy in terms of dislocation density-like variables which average over the discrete
dislocation configurations and represent the dislocation system on scales above the spacing of the individual 
dislocation lines. We study the simple case of two-dimensional systems of straight dislocation lines before we proceed to derive energy functionals for systems of three-dimensionally curved dislocation lines pertaining to a single, as well as to multiple slip systems. We then illustrate several applications of the theory including Debye screening of dislocations in two and three dimensions, and the derivation of back stress and friction stress terms entering the stress balance from the free energy functionals.  
\end{abstract}

\pacs{46.50.+a, 62.20.M-, 62.20.mm, 64.60.av}

\maketitle

\section{Introduction}

Any dislocation density based theory of dislocation dynamics under stress, and thus of plasticity, must of necessity consist of two parts. The first consists in a density based description of dislocation {\em kinematics}, i.e., of the way dislocations as curved and connected lines move in space. The second consists in a density based description of dislocation {\em energetics} which allows to derive, via the powerful mathematical tools of variational calculus, the driving forces for dislocation motion. If one wants to go beyond the standard concepts of linear irreversible thermodynamics, a closed theory may require a third ingredient which provides the, in general non-linear, connection between driving forces and dislocation fluxes. 

As to describing the kinematics of dislocations on scales above the dislocation spacing where individual dislocation lines can no longer be resolved, significant progress has been made in recent years. In particular, the important question how to correctly describe the coupled kinematics of 'statistically stored' and geometrically necessary dislocations for three-dimensional (3D) dislocation systems has been addressed in several works \cite{ElAzab2000,Arsenlis2004,Acharya2006a,Acharya2006b,Hochrainer2007,Hochrainer2014,Hochrainer2015}. The progress in dislocation kinematics calls for a matching effort to develop averaged, density based descriptions of the energetics of dislocation systems. Such an effort needs to consider both excess dislocations associated with the spatial average of the classical dislocation density tensor (often termed geometrically necessary dislocations) and so-called statistically stored dislocations of zero net Burgers vector which dominate plasticity in the early stages of deformation. 

For dislocation systems described as systems of discrete lines, expressions for the associated elastic energy functional and interaction stresses have been provided in the classical works of dislocation theory, see e.g \cite{deWit1960} and, for overwiew, the textbook of Hirth and Lothe \cite{Hirth1982}. Recent progress has focused on methods to regularize the elastic singularity and associated diverging energy density in the dislocation core, using either nonlocal elasticity theories (e.g. \cite{Lazar2005} or continuous Burgers vector distributions \cite{Cai2006}. Starting from the energy functional of a discrete dislocation system, one may then develop an energy functional for dislocation densities through appropriate averaging procedures. This is the approach pursued in the present investigation. 

There are alternative approaches. In the traditional spirit of constitutive modelling in continuum mechanics, numerous authors have introduced free energy functionals which depend on dislocation density-like variables where the related functional dependencies were assumed in an ad-hoc manner, see \cite{Gurtin2008} as one example of many. Usually, little attention is paid to the question how, if at all, these functionals can be derived from properties of the underlying discrete dislocation systems. We find this method of reasoning of little relevance to our own work, however, we note that our results can serve as a benchmark to assess whether or not the structure of various free energy functionals used in the literature is adequate for representing actual dislocation systems. Groma and co-workers derived equations for dislocation density evolution in 2D by direct averaging of the discrete dynamics \cite{Groma1997,Zaiser2001,Groma2003} and then proceeded to design, via educated guess, free energy functionals which are consistent with this averaged dynamics \cite{Groma2006}, or which are directly designed to reproduce average behavior in discrete dislocation dynamics simulations \cite{Groma2015}. In the present work we pursue a modified but related approach - instead of systematically averaging the dynamics, and then matching it with free energy functionals obtained via educated guess, we apply systematic averaging to directly obtain an energy function. 

Recently, several authors have attempted to derive free energy functionals for dislocation systems using thermodynamic formalisms, see e.g. \cite{Le2014,Kooiman2014,Kooiman2015}. In the opinion of the present author, thermodynamic approaches need to deal with the fundamental problem that dislocations do not exist in thermal equilibrium, and some authors may not be sufficiently aware of the implications of this basic fact. To illustrate the problem, let us compute the energy associated with a single Burgers vector of dislocation line length. This energy is of the order of $\mu b^3$ where $b$ is the length of the Burgers vector and $\mu$ is the shear modulus of the material. For typical materials parameters of copper, this energy amounts to about $4$ eV. To provide matching thermal energies one would need to consider  temperatures of the order of 40000 K, which is more than one order of magnitude above the melting temperature. Kooiman et. al. \cite{Kooiman2014} make the same observation when they notice that the coupling constant which gives the relative magnitude of elastic to thermal energies is, for dislocation systems at room temperature, of the order of 100. Hence, thermal effects and thus classical thermodynamic entropy are practically irrelevant to dislocation systems, whose driving forces derive almost exclusively from the internal (elastic) energy. Attempts to evaluate properties such as the range of correlations in dislocation systems using thermodynamic formalisms have produced interesting results, see e.g. the remarkable work of Limkumnerd and Van der Giessen \cite{Limkumnerd2008} who use a Langevin-type approach to evaluate correlation functions, with results that are consistent with the results of discrete dislocation dynamics (DDD) simulations. However, these authors recognize that the fluctuation magnitude that must be assumed to achieve such agreement is orders of magnitude above the level of thermal fluctuations even at the melting temperature, hence it represents some kind of "effective temperature" -- in fact a fit parameter that needs to be adjusted to make the range of dislocation correlations equal to a few dislocation spacings as observed in simulations. An effective temperature is also introduced by Groma et. al. \cite{Groma2006} in their free energy expression, and these authors make the same observation. These approaches may be considered implementations of the general suggestion by Berdichevsky  to consider microstructural disorder in terms of an effective microstructural entropy and associated temperature \cite{Berdichevsky2008}.

The bottomline is that, in order to explain why for instance the screening radius in a dislocation system is of the order of several dislocation spacings \cite{Ungar1984,Groma2006}, one needs to introduce effective temperatures that cannot be related to standard temperature and thus, effectively, constitute phenomenological fit parameters. 

So why do we find, in plastically deformed crystals, densities of 'statistically stored' dislocations (dislocations which have zero net Burgers vector and might thus annihilate) which are as high as 10$^{15}$m$^{-2}$ \cite{Ungar1984}, when equilibrium thermodynamics requires this density to be zero? The answer is simply that statistically stored dislocations exist because kinematic constraints prevent them from annihilating. One can understand the problem best by considering dipoles consisting of dislocations of opposite sign moving on parallel slip planes: As long as the interaction is not sufficient to overcome the high energy barrier that prevents dislocations from leaving their slip planes, the dislocations will form a dipole with a width that is dictated by the slip plane spacing. Furthermore, dislocations are likely to become trapped in the first local energy minimum close to their initial position, and thermal energies (which are orders of magnitude less than dislocation interaction energies) may be unable to liberate them. Metastability and kinematic constraints ensure that dislocations are, and tend to remain, in the crystal despite the fact that in  thermal equilibrium they should not be there. This raises major conceptual problems. Unless methods for dealing with metastability and constraints in statistical thermodynamics are developed to a much higher level than presently available, it may be difficult for us to derive the non-equilibrium statistical properties of dislocation systems from first principles. The present work therefore pursues more modest goals: To establish the fundamental structure of functionals which express the energy of a dislocation system in terms of dislocation density functions, to clearly formulate the parts of the functionals which we can know for certain, and to find reasonable approximations for those which depend on information regarding the relative arrangement of dislocations. In this task, we are inspired by density functional theory of electron systems where energy contributions of known and established form which represent long-range electrostatic interactions (the Hartree energy functional) are separated from those which depend on correlations in a many-body problem  (the exchange-correlation energy), see e.g. \cite{Evarestov2007} or other textbooks on quantum mechanics of many-electron systems . The latter are approximated by reference to idealized systems such as a homogeneous electron gas. We apply the same strategy to the many-dislocation problem. Just as in electron theory, what we get is the fundamental mathematical structure of the energy functional which we determine first in two and then in three dimensions. As to the correlation energies, these depend on parameters characterizing the range and nature of dislocation-dislocation correlations -- parameters which remain to be determined by reference to direct DDD simulations of the many-dislocation problem. Again, this strategy is analogous to the proceedings in density functional theory where exchange-correlation energy functionals are formulated and parameterized by reference to idealized model systems (thee free electron gas) or by comparison with direct numerical simulations of the many-body problem via quantum Monte Carlo methods. 

The technical method which we shall use is to represent the dislocation interaction energy in terms of densities and correlation functions. This idea is not new: Correlation functions have been introduced for averaging the forces in 2D dislocation systems, and hence the dynamics, in earlier work by Zaiser and Groma \cite{Zaiser2001,Groma2003}. Averaging the forces implies, of course, evaluating average derivatives of the elastic energy functional. Here we apply the same averaging method to evaluate the energy functional itself. A similar approach has been used already in work the 1960s, see e.g the work of Kocks and Scattergood \cite{Kocks1969} on systems of straight parallel dislocations, but was not further pursued. A possible reason for this lies in the fact that, in absence of DDD simulations which can provide complete information about the dislocation microstructure, information about dislocation correlations is hard to come by -- even though some limited information can be inferred from electron microscopy and X-ray profile analysis data, and indeed some of the early work by Wilkens in the field considers both mean square stresses (and thus elastic energy densities) and X-ray line broadening \cite{Wilkens1969,Wilkens1987}. In the present study we resume these approaches and extend them to general dislocation systems in three dimensions with multiple slip systems and arbitrarily curved dislocations. We first revisit results of classical dislocation theory for the energy of discrete dislocation systems, and then develop our averaging methodology for the conceptually simple case of systems of straight parallel edge dislocations. We generalize the results first to curved dislocations on a single slip system, and then to general 3D dislocation systems. We then demonstrate a few applications for the resulting free energy functionals, first to evaluate 'Debye screening' of dislocations in 2D, i.e., the formation of an induced distribution of local excess Burgers dislocations around a given dislocation which screens the long-range dislocation stress field. We also investigate the emergence of 'back stress' terms in the stress balance that are proportional to second-order in the plastic strain gradients and demonstrate that these are associated with energy contributions that are quadratic functionals of the local excess dislocation density. Finally, we outline how our results can be used to evaluate the 'friction stress' associated with formation and breaking of junctions in general 3D dislocation systems. We conclude with a brief discussion which puts our results into the context of other published work.

\section{Energy of a discrete dislocation system}

As demonstrated by de Wit \cite{deWit1960}, the energy of a three dimensional system of dislocations can be written in terms of double integrals over the dislocation lines. This representation has been directly implemented in discrete dislocation dynamics (DDD) codes, notably the parametric DDD model of Ghoniem and co-workers \cite{Ghoniem1999,Ghoniem2000} who use the variation of the energy with respect to dislocation line parameters in order to derive generalized forces acting on the dislocation lines. In our presentation we follow the representation given in the standard textbook of Hirth and Lothe \cite{Hirth1982} which can be applied both to closed loops and to loop segments. We write the energy of a dislocation system consisting of closed loops ${\cal C}^{(i)}$ with Burgers vectors $\Bb^{(i)} = b \Be^{(i)}$ as
\begin{equation}
E = \frac{1}{2}\sum_{ij} \oint_{{\cal C}^{(i)}} \oint_{{\cal C}^{(j)}}  \Bl^{(i)}.{\BE}^{(i,j)}(\Br^{(i)}-\Br^{(j)}).\Bl^{(j)}  \rmd s^{(i)} \rmd s^{(j)}.
\label{eq:deWit}
\end{equation}
Here the sum runs over all pairs of loops, and the self-energy of each loop is evaluated as half the interaction energy of two loops at distance $b$ (more generally, the core radius). Here and in the following, upper bracketed indices $^{(i)}$ enumerate dislocation loops or segments of loops, whereas lower indices indicate coordinates of a Cartesian coordinate system. When dealing with 3D dislocation networks, we retain the decomposition into closed loops but break the loops into segments ${\cal S}$ separated by nodes. A junction which forms at the intersection of two loops ${\cal C}^{(i)}$ and ${\cal C}^{(j)}$ is thus represented as two segments of Burgers vectors $\Bb^{(i)}$ and $\Bb^{(j)}$ that are aligned with each other between the two nodes which delimit the junction. A collinear reaction where $\Bb^{(i)} = - \Bb^{(j)}$ is represented as two aligned segments of opposite Burgers vector, not as a missing segment. In segment representation, the energy of the dislocation system is
\begin{equation}
E = \frac{1}{2}\sum_{ij} \int_{{\cal S}^{(i)}} \int_{{\cal S}^{(j)}}  \Bl^{(i)}.{\BE}^{(i,j)}(\Br^{(i)}-\Br^{(j)}).\Bl^{(j)}  \rmd s^{(i)} \rmd s^{(j)}.
\label{eq:deWitSeg}
\end{equation}

In Eqs. (\ref{eq:deWit}) and (\ref{eq:deWitSeg}), the interaction kernel ${\BE}^{(i,j)}$ is given by
\begin{equation}
{\cal E}^{(i,j)}(\Br^{(i)}-\Br^{(j)}) = \frac{\mu b^2}{4\pi} \Bg^{(i,j)}(\Br^{(i)}-\Br^{(j)}) 
\end{equation}
where 
\begin{eqnarray}
\Bg^{(i,j)}(\Br^{(i)}-\Br^{(j)}) = - \left[\Be^{(i)}\otimes \Be^{(j)} -\Be^{(j)}\otimes \Be^{(i)}\right]{\rm Tr} \BG 
&+& \frac{\Be^{(i)}\otimes\Be^{(j)}}{2} {\rm Tr} \BG + \frac{1}{1-\nu} \tilde{\BG}(\Be^{(i)},\Be^{(j)}).
\label{eq:g}
\end{eqnarray}
In this expression, $\mu$ is the shear modulus, $\nu$ is Poisson's number, $\BG$ is a tensor with components 
\begin{equation}
G_{kl}(\Br^{(i)}-\Br^{(j)}) = \frac{\partial^2}{\partial r_k \partial r_l} |\Br^{(i)}-\Br^{(j)}| \quad,\quad {\rm Tr}G = G_{kk} = \frac{2}{|\Br^{(i)}-\Br^{(j)}|},
\end{equation}
and $\tilde{\BG}(\Be^{(i)},\Be^{(j)})$ has the components
\begin{equation}
\tilde{\BG}_{kp} = b^{(i)}_l \epsilon_{lkm} G_{mn} \epsilon_{nop} b^{(j)}_o.
\end{equation}
In equation (\ref{eq:g}), the first term on the right-hand side is non-zero only if neither the line directions nor the Burgers vectors of both segments are aligned with each other, hence it can be considered to describe edge-screw interactions. The second term on the right-hand side describes the interactions of the screw components of both line segments, and the third term describes the interactions of edge components. 

These equations apply to dislocations in an infinite medium. In the presence of boundaries and boundary tractions which cause, in a fictitious crystal without dislocations, the stress field $\Bsigma_{\rm ext}(\Br)$, the energy changes. The corresponding energy contribution can be written in terms of the fictitious work that would need to be done by the Peach-Koehler forces in order to expand the loops ${\cal C}$ to their current size. For a system of planar glide loops this is simply given by
\begin{equation}
E_{\rm ext} = \sum_{i} \int_{{\cal A}^{(i)}} b \BMM^{(i)} : \Bsigma_{\rm ext}(\Br) \rmd^2 r.
\label{eq:Eext}
\end{equation}
Here ${\cal A}^{(i)}$ is the area enclosed by the loop ${\cal C}^{(i)}$ in the slip plane with normal $\Bn^{(i)}$ and the projection tensor $\BMM^{(i)} = (\Be^{(i)}\otimes \Bn^{(i)} + \Bn^{(i)} \otimes \Be^{(i)})/2$. Alternatively, we may write the same expression in terms of the microscopic plastic strain 
\begin{equation}
\BM{\varepsilon}^{\rm pl,d}(\Br) = \sum_{i} \int_{{\cal A}^{(i)}} b \BMM^{(i)} \delta(\Br - \Br') \rmd^2 r'.
\end{equation}
Inserting into Eq. (\ref{eq:Eext}) gives
\begin{equation}
E_{\rm ext} = \int_V \BM{\varepsilon}^{\rm pl,d}(\Br):\Bsigma_{\rm ext}(\Br) \rmd^3 r.
\label{eq:Eext2}
\end{equation}
where the integration is carried over the crystal volume. In the following it will be useful to develop a number of ideas first for the physically unrealistic, but conceptually simple case of quasi-two-dimensional (2D) systems consisting of straight parallel edge dislocations pertaining to a single slip system. For such a dislocation system, we may without loss of generality set $\Be^{(i)} = \Be_x$ and $\Bl^{(i)} = s^{(i)} \Be_z$ where $s^{(i)} \in [1,-1]$ is the sign of a dislocation. The line integrals then reduce to integrals over the $z$ axis, and the energy of the system becomes
\begin{equation}
E = \frac{1}{2}\sum_{i \neq j} s^{(i)} s^{(j)} E_{\rm int}(\Br^{(i)}-\Br^{(j)}) + \sum_i E_{\rm self}
\end{equation}
where the vectors $\Br^{(i)}$ now lie in the $xy$ plane and all energies are understood as energies per unit length in $z$ direction. The self and interaction energies are given by (see e.g. \cite{Hirth1982})
\begin{equation}
E_{\rm self} = \frac{\mu b^2 }{4\pi(1 - \nu)} \ln \left(\frac{R}{\alpha b}\right)\quad,\quad E_{\rm int}(x,y) = - \frac{\mu b^2}{2 \pi (1-\nu)}\left[\ln\left(\frac{r}{R}\right) + \frac{y^2}{r^2}\right] 
\end{equation}
where $R$ is the crystal radius (or another external dimension of the system) and $r = (x^2 + y^2)^{1/2}$ is the spacing of the dislocations in the $xy$ plane. $\alpha b$ is a measure of the dislocation core radius, with the parameter $\alpha \approx 1$ chosen to correctly represent the core energy of the dislocations.  Finally, the external energy for the considered dislocation system reads  
\begin{equation}
E_{\rm ext} = \sum_{i} s^{(i)} b \int_{x_0}^{x^{(i)}} \tau_{\rm ext}(x,y^{(i)}) \rmd x 
\label{eq:Eext2D}
\end{equation}
where $\tau_{\rm ext} = \BMM:\Bsigma_{\rm ext}$ with the projection tensor $\BM = (\Be_x \otimes \Be_z + \Be_z \otimes \Be_x)/2 $ is the resolved shear stress in the slip system. In Eq. (\ref
{eq:Eext2D}) the glide distance travelled by dislocation $^(i)$ ranges from an arbitrary reference position $x_0$ to its current position $x^(i)$. 

\section{Density functional theory of two-dimensional dislocation systems}

To write the energy of our 2D model system as a functional of the dislocation densities, we define discrete densities of dislocations of sign $s$, and discrete 
pair densities of dislocation pairs of signs $(s,s')$, as
\begin{equation}
\rho_s^{\rmd}(\Br) = \sum_{j: s^{(j)}=s} \delta(\Br - \Br^{(j)})\quad,\quad
\rho_{ss'}^{\rm d,p}(\Br,\Br') = \sum_{\substack{k: s^{(k)}=s' \\ j: s^{(j)}=s \\ j\neq k}} \delta(\Br - \Br^{(j)})\delta(\Br' - \Br^{(k)});.
\end{equation}
With these, the energy of the dislocation system can be written as
\begin{equation}
E = \sum_s \int \rho_s^{\rmd}(\Br) = E_{\rm self} \rmd^2 r 
+ \frac{1}{2}\sum_{ss'} s s' \iint E_{\rm int}(\Br -\Br') \rho^{\rm d,p}_{ss'}(\Br,\Br') \rmd^2 r \rmd^2 r' 
\end{equation}
where $N$ is the total number of dislocations. We now make a transition towards continuous densities via an averaging operation $\langle \dots \rangle$ (see Appendix A). This leads to
\begin{equation}
E = \sum_s \int \rho_s(\Br) E_{\rm self} \rmd^2 r +\frac{1}{2} \sum_{ss'} s s' \int \int E_{\rm int}(\Br -\Br') \rho_{ss'}(\Br,\Br') \rmd^2 r \rmd^2 r' \;.
\end{equation}
Here, the averaged single-particle densities $\rho_s(\Br) = \langle \rho^{\rmd}_{s}(\Br) \rangle$ can be understood as averages of the sign-dependent discrete densities, and the averaged pair densities $\rho_{ss'}(\Br,\Br') = \langle \rho^{\rmd}_{s}(\Br)\rho^{\rmd}_{s'}(\Br') \rangle$ are averages of products of discrete densities. Note that the averaged pair densities are in general {\em not} equal to the products of the averaged single-dislocation densities: averaging is a linear operation which does not interchange with the formation of a product. 

Hence, the information contained in the single-particle densities is of necessity incomplete. Nevertheless it is our goal to express the energy functional in terms of the densities $\rho_s(\Br)$. To this end we write the pair densities without loss of generality as 
\begin{equation}
\rho_{ss'}(\Br,\Br') = \rho_s(\Br)\rho_{s'}(\Br')[1 + d_{ss'}(\Br,\Br')] \;,
\end{equation}
where $d_{ss'}$ are correlation functions. This allows us to split the energy functional into a part which can be exactly expressed in terms of the dislocation densities ('Hartree Energy' $E_H$), and a part which depends on the correlation functions ('Correlation Energy' $E_C$) and needs to be evaluated in an approximate manner. After some algebra we arrive at 
\begin{eqnarray}
E &=& E_{\rm S} + E_{\rm H} + E_{\rm C}\nonumber\\
&=& \int \rho(\Br) E_{\rm self} \rmd^2 r + \frac{1}{2} \iint \kappa(\Br) \kappa(\Br') E_{\rm int}(\Br - \Br') \rmd^2 r \rmd^2 r'\nonumber\\
&+& \frac{1}{2} \sum_{ss'} ss' \iint \rho_s(\Br)\rho_{s'}(\Br')d_{ss'}(\Br,\Br')E_{\rm int}(\Br - \Br')  \rmd^2 r \rmd^2 r'\;.
\label{eq:Etot2D}
\end{eqnarray}
Here we have introduced the notations 
\begin{equation}
\rho(\Br) = \sum_s \rho_s(\Br)\quad,\quad \kappa(\Br) = \sum_s s \rho_s(\Br)
\label{eq:rhokappa}
\end{equation}
for the total and excess dislocation densities. 

\subsection{The Hartree or self-consistent energy}

The Hartree energy $E_{\rm H}$ depends only on the excess dislocation density. To analyze its meaning, we use that $\kappa = -(1/b) \partial_x \gamma$ where $\gamma$ is the mesoscopically averaged shear strain on the single slip system. We may now integrate the expression for $E_H$ twice by parts to write the Hartree energy as a functional of the mesoscopically averaged plastic strain $\BM{\varepsilon}^{\rm p} = \BMM{\gamma}^{\rm p}$:
\begin{eqnarray}
E_{\rm H} = \iint \BM{\varepsilon}^{\rm p}(\Br):\BM{\Gamma}(\Br - \Br'):\BM{\varepsilon}^{\rm p}(\Br')  \rmd^2 r \rmd^2 r'.
\label{eq:EHGF}
\end{eqnarray}
Here, for this particular problem, $\BM{\Gamma} = (\BMM^{-1} \otimes \BMM^{-1}) \partial^2 E_{\rm int}/\partial x^2$. We can re-write Eq. (\ref{eq:EHGF}) as
\begin{eqnarray}
E_{\rm H} = \int \BM{\varepsilon}^{\rm p}(\Br):\BM{\sigma}_{\rm int}(\Br)  \rmd^2 r
\end{eqnarray}
where the internal stress field is given by
\begin{eqnarray}
\BM{\sigma}_{\rm int}(\Br) = \int  \BM{\Gamma}(\Br - \Br'):{\varepsilon}^{\rm p}(\Br') \rmd^2 r'.
\end{eqnarray}
This corresponds to the solution of the elastic eigenstrain problem in an infinite medium by means of a Green's function method, see e.g. \cite{Zaiser2005} where expressions for $\BM{\Gamma}$ are given for the case of a general plastic strain field. Thus, the Hartree energy is just the elastic energy associated with the elastic-plastic problem in the absence of boundary effects. In the general case where boundaries are present, the boundary conditions result in an additional, 'external' stress field $\Bsigma_{\rm ext}$ which superimposes on the internal stress field $\Bsigma_{\rm int}$ and which enters into the energy $E_{\rm ext}$. In most practical circumstances where elastic-plastic problems are to be solved, the Hartree energy or its functional derivative (the internal stress) will not be computed from the dislocation field $\kappa$ but evaluated in conjunction with the 'external' stress from the solution of the elastic boundary value problem.  

In summary, the Hartree Energy represents the part of the elastic energy functional that is related to long-range internal stresses which can be described in terms of the coarse grained strain field $\BM{\varepsilon}^{\rm p}$ or its spatial derivative, the geometrically necessary dislocation (GND) density $\kappa$. In absence of mesoscale strain gradients ($\kappa = 0$, equal numbers of dislocations of both signs), this term is zero and, hence, all interaction energy terms are associated with correlations. We now focus on the correlation energy $E_{\rm C}$.

\subsection{The correlation energy}

To evaluate the correlation energy we proceed in the spirit of density functional theory of electron systems, i.e., we use a local density approximation where we approximate the correlation energy by a local functional of the dislocation densities which we evaluate as the correlation energy of a spatially homogeneous reference system. As shown in Appendix B, this is feasible if and only if the correlation functions $d_{ss'}(\Br,\Br')$ are short ranged functions of range $\ell$ which, for large values of $|\Br - \Br'| \gg {\cal \ell}$, go to zero faster than algebraically (short-range correlated/macro-disordered dislocation systems). The assumption that the $d_{ss'}$ are short ranged functions allows us, for dislocation arrangements where the densities $\rho_s$ are weakly space dependent on scale ${\ell}$, to approximate  
\begin{eqnarray}
 \rho_s(\vec{r}_1)\rho_{s'}(\vec{r}_2)d_{ss'}(\vec{ r}_1,\vec{r}_{2}) \approx 
 \rho_s(\vec{r}_1)\rho_{s'}(\vec{r}_1)d_{ss'}(\vec{ r}_1-\vec{r}_{2}).
\label{eq:drho2} 
\end{eqnarray}
This local density approximation represents the zeroth order of a systematic expansion which expresses the energy functional in terms of gradients of the dislocation densities of increasing order (see Appendix). From Eqs. (\ref{eq:Etot2D}) and (\ref{eq:drho2}) the correlation energy reads 
\begin{eqnarray}
E_{\rm C}= \frac{1}{2} \sum_{ss'} ss' \int\left[\rho_s(\vec{r})\rho_{s'}(\vec{r})\int E_{\rm int}(\vec{ r}')d_{ss'}(\vec{ r}') \rmd^2 r'\right]
\rmd^2 r\;.
\end{eqnarray}
Using the notation 
\begin{eqnarray}
E_{\rm int}(\Br) &=& \frac{\mu b^2}{2 \pi(1-\nu)} g(\Br) \quad,\quad g(\Br) = - \log\left(\frac{r}{R}\right) - \left(\frac{y}{r}\right)^2\,, 
\label{eq:e} 
\end{eqnarray}
this can be rewritten as
\begin{eqnarray}
E_{\rm C}= \frac{\mu b^2}{4 \pi(1-\nu)}\int \left [\sum_{ss'=\pm 1} 
\rho_s(\vec{r})\rho_{s'}(\vec{r}) T_{ss'}\right]{\rm d}^2r  \label{eq:Ecorr}
 \end{eqnarray}
where 
\begin{eqnarray}
 T_{ss'}=\int ss'd_{ss'}(\Br) g(\Br) \rmd^2 r. \label{eq:t}
\end{eqnarray}

To proceed further we need to specify some properties of the functions $d_{ss'}$. By construction (see Appendix A) the correlation functions have the properties
\begin{eqnarray}
\int d_{ss'}(\Br -\Br'){\rm d}^2 r' = 0, \ \ { \rm if}  \ \ s\neq s' \label{eq:dpm} \quad,
\int d_{ss}(\Br -\Br'){\rm d}^2 r' \approx -1/\rho_s(\Br) \ \ { \rm if}  \ \ s=s' . \label{eq:dppm}
\end{eqnarray}
where we have used that, in a weakly heterogeneous dislocation arrangement, $\rho_s(\Br)\approx \rho_s(\Br')$ changes little over the range of the function $d_{ss}$. (This weak heterogeneity condition is a requirement for the use of a local density approximation, see the appendix for an outline towards more general non-local approaches). Furthermore, as discussed in detail elsewhere \cite{Zaiser2001,Zaiser2014}, due to the scale free nature of dislocation-dislocation interactions in nearly homogeneous dislocation systems, any correlation functions which emerge spontaneously from the evolution of an initially disordered dislocation system must exhibit a range $\ell \propto \rho^{-1/2}$ that is proportional to the mean dislocation spacing, i.e., to the inverse square root of the total dislocation density as defined by equation (\ref{eq:rhokappa}).  Hence we assume that the correlation functions depend only on the relative position of the two dislocations divided by the mean dislocation spacing, expressed through the variable $\Bu = (\Br-\Br')\sqrt{\rho}$, {\it i.e.} $d_{ss'}(\Br - \Br') = d_{ss'}(\Bu)$. Note that, since $\rho_s(\Br)\approx \rho_s(\Br')$ over the range of the correlation function $d_{ss'}$ , it does not matter whether we evaluate $\rho$ at $\Br$ or at $\Br'$.

Let us now first consider the case of $s\neq s'$. As the first step the radial function $g$ given by Eq. (\ref{eq:e}) 
is rewritten as
\begin{eqnarray}
g(\Br) = g_0 + g_r(r \sqrt{\rho}) \quad{\rm where} \quad g_0 = \ln\left(R\sqrt{\rho}\right) \;,\; g_r = - \ln\left(r \sqrt{\rho} \right) - \frac{y^2}{r^2};.
\label{eq:Vr3}
\end{eqnarray}
After substituting Eq. (\ref{eq:Vr3}) into Eq. (\ref{eq:t}) and using the condition (\ref{eq:dpm}) one obtains that
\begin{eqnarray}
 T_{ss'} &=& - \int d_{ss'}(\Br) g_r(\Br)\rmd^2 r\nonumber\\
 &=& \frac{1}{\rho} \int d_{ss'}(\Bu)\left[\ln(u) + \frac{u_y^2}{u^2}\right] d^2 u = \frac{D_{ss'}}{\rho}. \label{eq:tpm}
\end{eqnarray}
It can be seen that the correlation function enters our further considerations only in form of the dimensionless number $D_{ss'}$. 

In the case $s=s'$, we substitute Eq. (\ref{eq:Vr3}) into Eq. (\ref{eq:t}). Due to Eq. (\ref{eq:dppm}) we find that
\begin{eqnarray}
T_{ss}&=&\int d_{ss}(\Br) g_{r}(\Br)\rmd^2 r -\frac{1}{2\rho_s}\ln\left(\rho R^2 \right) \nonumber\\
&=& \frac{1}{\rho} \int d_{ss}(\Bu)\left[- \ln(u) - \frac{u_y^2}{u^2}\right] d^2 u  - \frac{1}{2\rho_s}\ln\left(\rho R^2 \right) \nonumber\\
&=& \frac{D_{ss}}{\rho} - \frac{1}{2\rho_s}\ln\left(\rho_s R^2 \right).
\label{eq:tpp}
\end{eqnarray}
By substituting Eqs. (\ref{eq:tpm},\ref{eq:tpp}) into Eq. (\ref{eq:Ecorr}) we get 
\begin{eqnarray}
 E_{\rm C} &=& \frac{\mu b^2}{4 \pi(1-\nu)} \int \left[(D_{+-}+D_{-+})\frac{\rho_+\rho_-}{\rho} + D_{++} \frac{\rho_+^2}{\rho} + D_{--} \frac{\rho_-^2}{\rho}  - \frac{\rho_++\rho_-}{2} \ln(\rho R^2) \right] \rmd^2 r\nonumber\\
&=& \frac{\mu b^2}{4 \pi(1-\nu)} \int \left[ \rho \left(\frac{D_{\rm I}}{4} - \frac {1}{2} \ln(\rho R^2)\right) + \frac{D_{\rm II}}{4} \frac{\kappa^2}{\rho} \right]]\rmd^2 r = \frac{\mu b^2}{4 \pi(1-\nu)} \int \left[ - \frac{\rho}{2} \ln \left(\frac{\rho R^2}{a^2}\right) + \frac{D_{\rm II}}{4} \frac{\kappa^2}{\rho} \right]\rmd^2 r\;,
\label{eq:ecor}
\end{eqnarray}
where $D_{\rm I} = \sum_{ss'} ss' D_{ss'}$ and $D_{\rm II} = \sum_{ss'} D_{ss'}$. Here we have used that $\rho_{\pm} = (\rho \pm \kappa)/2$ and, in a weakly polarized dislocation arrangement, $D_{++} = D_{--}$. The non-dimensional parameter $a = \exp(D_{\rm I}/2)$ can be envisaged as a dislocation screening radius, measured in units of mean dislocation spacings. We see that the correlation energy in local density approximation is indeed a local functional of the dislocation density functions. It consists of two contributions: first, the term proportional to $\rho$ can be envisaged as a screening energy which reduces the dislocation energy as compared to a random dislocation arrangement. Second, the term quadratic in $\kappa$ characterizes modifications to local screening as we move from an unpolarized to polarized dislocation arrangements.

The total energy function of the dislocation arrangement then reads 
\begin{eqnarray}
E = \frac{\mu b^2}{4\pi(1-\nu)}\left[- \int \rho \ln \left(\frac{\rho}{\rho_0}\right) + \int \frac{D_{\rm II}}{4} \frac{\kappa^2}{\rho} +
\frac{1}{2} \iint \kappa(\Br) \kappa(\Br')g(\Br - \Br') \rmd^2 r'\right] \rmd^2 r'\;.
\label{eq:Etot2Dfinal}
\end{eqnarray}
Here the normalization constant $\rho_0 \gg \rho$ is given by $\rho_0 = a^2/(\alpha^2 b^2)$. The total energy consists of three contributions: The term proportional to $\rho$ can be envisaged as line energy of a dislocation which is screened by the surrounding dislocations. The quadratic but local term in $\kappa$ is a correction to screening, and finally, the Hartree Energy expressed by the double integral over $\kappa$ describes the energy stored in the long-range elastic field associated with macroscopic polarization of the dislocation arrangement.

\section{Density functional theory of three-dimensional dislocation systems}

\subsection{System of loops on a single slip system}

We first consider a 3D system of loops pertaining to a single slip system with slip vector $\Be^{(i)}=\Be_x$ and slip plane normal $\Bn = \Be_y$. To facilitate the transfer of our results to the general case of multiple slip systems, we start from the `segment representation' of the dislocation energy, Eq. (\ref{eq:deWitSeg}). For loops on a single slip system, there exists no natural subdivision into segments as provided by the nodes in a 3D dislocation network. Instead, the length of the segments ${\cal S}$ is an artificial parameter of the calculation, which will be chosen as a small fraction $\eta$ of the local radius of curvature $R_{\rm c}$ of the dislocation lines -- a method also used in discrete dislocation schemes which represent the dislocation as a sequence of straight segments separated by nodes \cite{Weygand2002}. We split the elastic energy of the system, Eq. (\ref{eq:deWitSeg}), into sums of segment self energies and segment interaction energies: 
\begin{eqnarray}
E &=& E_{\rm S} + E_{\rm I}\nonumber\\
&=& \frac{1}{2}\sum_{i} \int_{{\cal S}^{(i)}} \int_{{\cal S}^{(i)}}  \Bl(s).{\BE}(\Br(s)-\Br(s')).\Bl(s')  \rmd s \rmd s'\nonumber\\
&+& \frac{1}{2}\sum_{i\neq j} \int_{{\cal S}^{(i)}} \int_{{\cal S}^{(j)}}  \Bl^{(i)}.{\BE}(\Br^{(i)}-\Br^{(j)}).\Bl^{(j)}  \rmd s^{(i)} \rmd s^{(j)}.
\label{eq:deWitSelfInt}
\end{eqnarray}
Here, $\BE$ is the interaction energy tensor for segments in the considered slip system which we write as 
\begin{equation}
\BE(\Br-\Br') = \frac{\mu b}{4 \pi} \Bg(\Br-\Br')
\end{equation}
where $\Bg$ follows from Eq. (\ref{eq:g}) with $\Be_1 = \Be_2 = \Be_x$. Since we consider glide loops on a single slip system where the line direction $\Bl$ is contained in the plane $y=0$, this tensor has only two relevant components which are explicitly given by
\begin{equation}
g_{xx}(\Br - \Br') = \frac{1}{|\Br-\Br'|} \quad, \quad g_{zz}(\Br - \Br') = \frac{1}{1-\nu}
\frac{\partial^2}{\partial y^2}|\Br -\Br'|.
\end{equation}
In a first approximation, the segment self energies are replaced by the self energies of straight segments which we write, following Hirth and Lothe \cite{Hirth1982}, as
\begin{equation}
E_{\rm S} \approx \sum_{i} \int_{{\cal S}^{(i)}_{L}}  \Bl(s).{\BE}_L.\Bl(s)  \rmd s 
\label{eq:ESelf}
\end{equation}
where the line energy tensor is given by
\begin{equation}
{\BE}_L = \frac{\mu b^2}{4 \pi}{\Bg_L} \quad,\quad \Bg_L = \frac{\left(\BI - \nu \Be_x\otimes \Be_x \right)}{1-\nu} \ln\left(\frac{L}{b}\right).
\label{eq:ELine}
\end{equation}
In Eq. (\ref{eq:ESelf}), ${\cal S}^{(i)}_{L}$ denotes a straight segment of length $L$ connecting the endpoints of segment ${\cal S}^{(i)}$. In the spirit of nodal discrete dislocation dynamics \cite{Weygand2002}, we take the segment length to be a small fraction of the radius of curvature of the dislocation line under consideration, $L = \eta/k$ where the dislocation curvature $k$ characterizes the line shape on scales above the segment length. The deviation between the curved segment and its straight approximation goes to zero in proportion with $\eta^2$ as $\eta \to 0$. In practice, $\eta$ may be adjusted to provide an optimum representation of the core energy contribution to the dislocation self energy.

We now proceed in direct generalization of the two-dimensional case. We define the discrete dislocation segment density as 
\begin{equation}
\rho_{\rm d}(\Br,\phi) = \sum_{\cal S} \int_{\cal S} \delta(\Br - \Br(s)) \delta(\phi - \phi(s)) \rmd s\;,
\end{equation}
where $\phi$  is the angle between the Burgers vector $\Bb$ and the line direction $\Bl$. Similarly, we define the segment pair density as
\begin{equation}
\rho_{\rm d}^{\rm p}(\Br,\phi,\Br',\phi') = \sum_{{\cal S'}\neq{\cal S}} \iint_{\cal SS'} \delta(\Br - \Br(s)) \delta(\Br' - \Br(s')) \delta(\phi - \phi(s))\delta(\phi' - \phi(s')) \rmd s \rmd s'
\end{equation}
This allows us to write the dislocation self energy as 
\begin{eqnarray}
E_{\rm S} = \iint  \Bl(\phi). {\BE}_L .\Bl(\phi) \rho_{\rm d}(\Br,\phi) \rmd^3 r \rmd \phi
\label{eq:ESelf2}
\end{eqnarray}
and the dislocation interaction energy as
\begin{eqnarray}
E_{\rm I} =  \frac{1}{2}\iiiint \left[\Bl(\phi).{\BE}(\Br -\Br').\Bl(\phi')\right]\; \rho_{\rm d}^{\rm p}(\Br,\Br',\phi,\phi') \;\rmd^3r \rmd \phi \rmd^3 r' \rmd \phi' \;.
\label{eq:deWitD}
\end{eqnarray}

Upon averaging, the discrete densities and pair densities become continuous functions of their arguments. In Eq. (\ref{eq:ESelf2}), the only change is that we replace the discrete density $\rho_{\rm d}(\Br,\phi)$ by its continuous ensemble average $\rho(\Br,\phi)$. To evaluate the dislocation interaction energy, we proceed in analogy with the 2D case and write the ensemble averaged pair density as
\begin{equation}
\rho_{\rm p}(\Br,\phi,\Br',\phi') = \rho(\Br,\phi)\rho(\Br',\phi')[1 + d(\Br,\Br',\phi,\phi')] \;.
\label{eq:3dcor}
\end{equation}
Comparison demonstrates that the angle $\phi$ in the present formalism plays very much the same role as the 'sign' $s$ in the 2D problem. Inserting Eq. (\ref{eq:3dcor}) into Eq. (\ref{eq:deWitD}) allows us to separate the dislocation interaction energy $E_{\rm I} = E_{\rm H} + E_{\rm C}$ into Hartree and correlation energy terms. The self energy, Hartree energy and correlation energy are given by
\begin{eqnarray}
E_{\rm S} &=& \iint  \rho(\Br,\phi) [\Bl(\phi). {\BE}_L.\Bl(\phi)] \rmd^3 r \rmd \phi
\label{eq:ES}\\
E_{\rm H}
&=& \frac{1}{2} \iint \rho(\Br,\phi) \left[\Bl(\phi).{\BE}(\Br -\Br').\Bl(\phi')\right]\; \rho(\Br',\phi') \;\rmd \phi \rmd \phi' \;\rmd^3 r \rmd^3 r' 
\label{eq:EH}\\
E_{\rm C} &=& \frac{1}{2} \iint \rho(\Br,\phi) \left[\Bl(\phi).{\BE}(\Br -\Br').\Bl(\phi')\right]\; \rho(\Br',\phi') d(\Br,\Br',\phi,\phi') \;\rmd \phi \rmd \phi' \;\rmd^3 r \rmd^3 r'
\label{eq:EC}
\end{eqnarray}
We can eliminate the explicit angular dependencies from these equations by choosing an appropriate representation of the dislocation density functions. To this end we resort to the concept of so-called alignment tensors. 

\subsubsection{Alignment tensor representation of the dislocation density functions} 

To simplify the expressions it is convenient to use an idea of Hochrainer \cite{Hochrainer2015} and represent the angular dependence of the dislocation density function $\rho(\Br,\phi)$ in terms of an alignment tensor expansion. Following Hochrainer, we define the sequence of (reducible) dislocation density alignment tensors as  
\begin{eqnarray}
\Brho^{[0]}(\Br) &=& \int \rho(\Br,\phi) \rmd \phi =: \rho(\Br),\nonumber\\
\Brho^{[1]}(\Br) &=& \int \rho(\Br,\phi) \Bl(\phi) \rmd \phi =: \Bkappa(\Br), \nonumber\\
\Brho^{[2]}(\Br) &=& \int \rho(\Br,\phi) \Bl(\phi)\otimes\Bl(\phi) \rmd \phi, \nonumber\\
&& \dots\nonumber\\
\Brho^{[n]}(\Br) &=& \int \rho(\Br,\phi) \Bl(\phi) [\otimes \Bl(\phi)]^{n-1} \rmd \phi. 
\label{eq:rhoalign}
\end{eqnarray}
where $[\otimes \Bl]^n$ denotes a $n$-fold tensor product with the line direction vector $\Bl$. The zeroth-order alignment tensor is the conventional dislocation density (line length per unit volume). The first-order tensor (or dislocation density vector) has as its components the edge and screw contributions of the geometrically necessary dislocation density. The second-order tensor contains the information about the distribution of the total dislocation density over edge and screw orientations, and so on \cite{Hochrainer2015}. 

The dislocation density function $\rho(\Br,\phi)$ can be recovered from the alignment tensors as follows \cite{Hochrainer2015}: We denote the irreducible part of the tensor $\Brho^{[n]}$ as $\tilde{\Brho}^{[n]}$ with components $\tilde{\Brho}_{i_1\dots i_n}$. Furthermore, we denote as
$\tilde{\rho}^{[n]}(\phi)$ the $n$-fold contraction of $\tilde{\Brho}^{[n]}$ with the direction vector $\Bl(\phi)$. Then, 
\begin{equation}
\rho(\Br,\phi) = \frac{1}{2 \pi} \left( \rho^{[0]}(\Br) + \sum_n 2^n \tilde{\rho}^{[n]}(\Br,\phi) \right) \quad,\quad
\tilde{\rho}^{[n]}(\Br,\phi) = \tilde{\Brho}^{[n]}_{i_1\dots i_n}(\Br) l_{i_1}(\phi) \dots l_{i_n}(\phi)\;.
\label{eq:rhorep}
\end{equation}
 
The double angular dependency of the pair correlation function $d(\Br,\Br',\phi,\phi')$ can be expressed in terms of a double alignment tensor expansion. We define
\begin{equation}
\Bd^{[n,m]}(\Br,\Br') = \int d(\Br,\Br',\phi,\phi') \Bl(\phi) [\otimes \Bl(\phi)]^{n-1} \otimes \Bl(\phi')[\otimes \Bl(\phi')]^{m-1} \rmd \phi \rmd \phi'. 
\label{eq:dalign}
\end{equation}
This is an expansion on the direct product of the unit circle with itself as the $\Bl$ are unit vectors in the $xy$ plane. In terms of the associated irreducible tensors $\tilde{\Bd}^{[n,m]}$ the function $d$ can be represented as
\begin{equation}
d(\Br,\Br',\phi,\phi') = \frac{1}{4 \pi^2} \sum_{n,m=1}^{\infty} 2^{n+m} \tilde{d}^{[n,m]}(\Br,\Br',\phi,\phi') \quad,
\label{eq:drep}
\end{equation}
with the expansion coefficients
\begin{equation}
\tilde{d}^{[n,m]}(\Br,\Br',\phi,\phi') = \tilde{d}^{[n,m]}_{i_1\dots i_n,j_1\dots j_m}(\Br,\Br') 
l_{i_1}(\phi) \dots l_{i_n}(\phi)\; l_{j_1}(\phi') \dots l_{j_m}(\phi')\;.
\label{eq:dcoeff}
\end{equation}

\subsubsection{The self energy}

The different contributions to the energy of a dislocation system can be expressed in a natural manner in terms of the dislocation density alignment tensors. As immediately seen from Eq. (\ref{eq:EH}) and the definition of the second order dislocation density alignment tensor, the self energy of a dislocation system can be represented as 
\begin{equation}
E_{\rm S} = \int  \BE_L:\Brho^{[2]}(\Br) \rmd^3 r = \frac{\mu b^2}{4 \pi} \int  \Bg_L:\Brho^{[2]}(\Br) \rmd^3 r 
\label{eq:ESAlign}
\end{equation}
The term under the integral has the character of a local energy density which is evaluated as a double contraction of the line energy and second-order dislocation density alignment tensors. It depends both on the local density of dislocations and on their character (edge/screw).

\subsubsection{The Hartree energy}

According to Eq. (\ref{eq:EH}) and the definition of the first order dislocation density alignment tensor $\Brho^{[1]}$ (also termed dislocation density vector $\Bkappa$), the Hartree energy can be represented as 
\begin{equation}
E_{\rm H} = \frac{1}{2} \iint \Brho^{[1]}(\Br).\BE(\Br -\Br').\Brho^{[1]}(\Br') \rmd^3 r \rmd^3 r' 
\label{eq:EHAlign}\\
\end{equation}
Alternatively, we might express this energy in terms of the classical dislocation density tensor $\Balpha$ (the curl of the plastic distortion). To this end we note that $\Balpha = \Brho^{[1]} \otimes \Bb = \Bkappa \otimes \Bb$ and define the fourth-rank interaction energy tensor $\BR = \Bb\otimes \BE \otimes \Bb$ to write
\begin{equation}
E_{\rm H} = \frac{1}{2} \iint \Balpha(\Br).\BR(\Br -\Br').\Balpha'(\Br') \rmd^3 r \rmd^3 r' \;.
\label{eq:EHBerd}
\end{equation}
This is the formulation used by Berdichevsky \cite{Berdichevsky2006}.

\subsubsection{The correlation energy}

To evaluate the correlation energy $E_{\rm C}$, we make the same crucial approximation as in the case of the 2D dislocation system: We use a local density approximation based upon the idea that the correlation function $d$ is short ranged, and that we may approximate the correlation energy by the correlation energy of a homogeneous system. In lowest-order approximation we thus set
\begin{eqnarray}
\rho(\Br,\phi)\rho(\Br',\phi')d(\Br,\Br',\phi,\phi') &\approx& \rho(\Br,\phi)\rho(\Br,\phi')d(\Br-\Br',\phi,\phi') \nonumber\\
&=& \rho(\Br,\phi)\rho(\Br,\phi')  \sum_{m,n=0}^{\infty} \frac{2^{n+m}}{4\pi^2} \tilde{d}^{[m,n]}(\Br-\Br',\phi,\phi')\;.
\label{eq:3dcorapp}
\end{eqnarray}
We now insert this approximation into the correlation energy, Eq. (\ref{eq:EC}). The form of the expansion coefficients given by Eq. (\ref{eq:dcoeff}) then leads us to define a sequence of coupling tensors $\BT^{[n,m]}$ with components
\begin{equation}
T^{[n,m]}_{i_1\dots i_n,j_1\dots j_m} = \frac{2^{m+n-2}}{4\pi^2} \int \tilde{d}^{[n-1,m-1]}_{i_1\dots i_{n-1}, j_1 \dots j_{n-1}}(\Br)
g_{i_n,j_m}(\Br) \rmd^3 r,
\label{eq:ECoeffT}
\end{equation}
where we again have assumed short-ranged correlation functions to ensure existence of the spatial integrals. 
We find that the correlation energy can be written in terms of these coupling tensors as
\begin{eqnarray}
E_{\rm C} &=& \frac{\mu b^2}{8\pi} \sum_{n,m=1}^{\infty} \iint \rho(\Br,\phi) l_{i_1}(\phi)\dots l_{i_n}(\phi) T^{[n,m]}_{i_1\dots i_n,j_1\dots j_m} \rho(\Br',\phi')l_{j_1}(\phi')\dots l_{j_m}(\phi') \rmd \phi \rmd \phi'  \rmd^3 r\nonumber\\
&=& \frac{\mu b^2}{8\pi} \sum_{n,m=1}^{\infty} \iint \rho^{[n]}(\Br)  \stackrel{(n)}{:} \BT^{[n,m]} \stackrel{(m)}{:} \rho^{[m]}(\Br) \rmd^3 r 
\label{eq:ECLOC}
\end{eqnarray}
where $\stackrel{(n)}{:}$ denotes a $n$-fold contraction. We note that, for reasons of symmetry, all interaction tensors $\BT^{[n,m]}$ must vanish where $n+m$ is an odd number. Such tensors involve, in Eq. \ref{eq:ECLOC}, odd numbers of products of the director $\Bl$. Since $\Bl$ changes sign under coordinate inversion ($\Br \to -\Br$), whereas the energy does not, all terms in Eq. (\ref{eq:ECLOC}) with odd numbers of products of $\Bl$ must be zero.  

In case where dislocation correlations emerge from the evolution of a otherwise scale free dislocation system we expect them to obey the relation $d = d(\Br \sqrt{\rho}) =: \Bu$. Furthermore we note that $\Bg(\Br-\Br') = \sqrt{\rho} \Bg(\Bu - \Bu')$. Using these relations we can write the coupling tensors in analogy with Eq. (\ref{eq:tpm}) as 
\begin{equation}
\BT^{[n,m]} = \frac{\BD^{[n,m]}}{\rho} \quad,\quad 
D^{[n,m]}_{i_1\dots i_n,j_1\dots j_m} = \frac{2^{m+n-2}}{4\pi^2} \int \tilde{d}^{[n-1,m-1]}_{i_1\dots i_{n-1}, j_1 \dots j_{n-1}}(\Bu)
g_{i_n,j_m}(\Bu) \rmd^3 u. 
\label{eq:ECoeffD}
\end{equation}
where $\Bu = \Br \sqrt{\rho}$. 

\subsubsection{Energy functionals for continuum dislocation dynamics}

In practical terms, it is desirable to truncate the alignment tensor expansion at some low order. This is done in continuum dislocation dynamics theories which represent dislocation systems in terms of the evolution of dislocation density alignment tensors, restricting themselves to the alignment tensors of order zero and one \cite{Hochrainer2014} or orders one and two \cite{Hochrainer2015}. By evaluating the interaction coefficients $D^{[n,m]}$ using data from discrete dislocation dynamics simulations, we might arrive at unbiased estimates to which degree such truncated expansions faithfully represent the energetics of dislocation systems. 

We explicitly give the energy functional for an expansion containing the first and second order dislocation alignment tensors. The diagonal components of the second order alignment tensor $\rho_{\rm s} := \rho_{xx}^{[2]}$ and $\rho_{\rm e} := \rho_{yy}^{[2]}$ correspond to the screw and edge dislocation densities. Furthermore we use that, because of invariance under the transformation $(\phi \to -\phi, \phi' \to - \phi')$, the correlation alignment tensor $\Bd^{[1,1]}$ is diagonal with only non-vanishing components $d^{[1,1]}_{xx}$ and $d^{[1,1]}_{yy}$. The same is true for the interaction energy tensor with the only non-vanishing components $g_{xx}$ and $g_{yy}$.  With these notations we write
\begin{eqnarray}
E&=&E_{\rm S}+E_{\rm H} + E_{\rm C} \nonumber\\
&=& \frac{\mu b^2}{4\pi} \int \left[\rho_{\rm s}(\Br) + \frac{1}{1-\nu}\rho_{\rm e}\right]\ln\left(\frac{\rho(\Br)}{\rho_0(\Br)}\right) \rmd^3 r + \frac{\mu b^2}{8\pi} \iint \Bkappa(\Br). \Bg(\Br-\Br').\Bkappa(\Br') \rmd^3 r \rmd^3 r'\nonumber\\
&+& \frac{\mu b^2}{8\pi} \int \frac{\rho_{\rm s}^2 D_{\rm ss} + \rho_{\rm e}^2 D_{\rm ee} + D_{\rm se} \rho_{\rm s}\rho_{\rm e}}{\rho(\Br)} \rmd^3 r
+ \frac{\mu b^2}{8\pi} \int  \frac{\Bkappa(\Br). \BD^{[1,1]}.\Bkappa(\Br)}{\rho(\Br)} \rmd^3 r.
\label{eq:ECDD2}
\end{eqnarray}
The interaction coefficients for the dislocation densities are given by
\begin{eqnarray}
D_{\rm ss} &=&  \frac{1}{\pi^2}\int g_{xx}(\Bu) d_{xx}^{[1,1]}(\Bu) d^3 u = \frac{1}{\pi^2}\int g_{xx}(\Bu) d(\Bu,\phi,\phi')\cos\phi \cos \phi' \rmd^3 u \rmd \phi \rmd \phi'\nonumber\\
D_{\rm ee} &=&  \frac{1}{\pi^2}\int g_{zz}(\Bu) d_{zz}^{[1,1]}(\Bu) d^3 u =  \frac{1}{\pi^2}\int g_{zz}(\Bu) d(\Bu,\phi,\phi')\sin\phi \sin \phi' \rmd^3 u \rmd \phi \rmd \phi'\nonumber\\
D_{\rm se} &=&  \frac{1}{\pi^2}\int [g_{zz}(\Bu) d_{xx}^{[1,1]}(\Bu) + g_{xx}(\Bu) d_{zz}^{[1,1]}(\Bu)] d^3 u\nonumber\\
&&  =  \frac{1}{\pi^2}\int  d(\Bu,\phi,\phi') [g_{yy}(\Bu)\cos\phi\cos\phi' + g_{xx}(\Bu)\sin\phi \sin \phi'] \rmd^3 u \rmd \phi \rmd \phi
\end{eqnarray}
and the interaction matrix associated with the dislocation density vector $\Brho^{[1,1]} = \Bkappa$ is
\begin{eqnarray}
\BD^{[1,1]}&=& \frac{1}{4\pi^2} \int \Bg(\Bu) d^{[0,0]}(\Bu) \rmd^3 u = \frac{1}{4 \pi^2}\int \Bg(\Bu) d(\Bu,\phi,\phi') \rmd^3 u \rmd \phi \rmd \phi'.
\end{eqnarray}
The quantity $\rho_0(\Br) = q(\Br) \eta b$ in Eq. (\ref{eq:ECDD2}) relates to the so-called curvature density (a product of dislocation density and curvature) which is one of the field variables of continuum dislocation dynamics as introduced in \cite{Hochrainer2014}.

A simplified theory as proposed in \cite{Hochrainer2014} might only consider the dislocation density alignment tensors of order zero and one, hence retains only information about the total dislocation density $\rho$ and geometrically necessary dislocation density vector $\Bkappa$. Since no additional information is available, the alignment tensor of order two is then represented as 
$\Brho^{[2]}= (\rho/2) \BI$. With this simplification the energy functional becomes
\begin{eqnarray}
E&=& \frac{\mu b^2 (2-\nu)}{8\pi(1-\nu)} \int \rho(\Br) \ln\left(\frac{\rho(\Br)}{\rho_0^*(\Br)}\right)\rmd^3 r 
+ \frac{\mu b^2}{8\pi} \iint \Bkappa(\Br). \Bg(\Br-\Br').\Bkappa(\Br') \rmd^3 r \rmd^3 r'\nonumber\\
&+& \frac{\mu b^2}{8\pi} \int  \frac{\Bkappa(\Br). \BD^{[1,1]}.\Bkappa(\Br)}{\rho(\Br)} \rmd^3 r
\end{eqnarray}
Here, all coefficients of energy contributions proportional to $\rho$ have been absorbed into the scaling factor $\rho_0^* = \alpha q b$, where the numerical parameter $\alpha$ is proportional to $\exp[(D_{\rm ss} +D_{\rm ee} + D_{\rm se})/4]$.  

\subsection{Multiple slip systems}

In case of multiple slip systems, dislocations are likely to form 3D networks. We now consider as segments ${\cal S}^{i,\beta}$ stretches of dislocations bounded by two nodes in the dislocation network. The superscript $\beta$ distinguishes the different slip systems with Burgers vectors $b \Be^{\beta}$, slip plane normal vectors $\Bn^{\beta}$ and projection tensors $\BMM^{\beta}$. The interactions between segments pertaining to two slip systems $\beta$ and $\beta'$ are given by ${\BE}^{\beta\beta'} = \BE^{(ij)}$ where $\Be^{(i)}= \Be^{\beta},\Be^{(j)}= \Be^{\beta'}$. For self energies we again approximate the self energy of a segment by that of a straight segment, for which we introduce the line energy tensor 
\begin{equation}
\BE_{L}^{\beta}  = - \frac{\mu b^2}{4 \pi}{\Bg_L^{\beta}} \quad,\quad \Bg_L = \frac{1}{1-\nu}\left(\BI - \nu \Be^{\beta}\otimes \Be^{\beta} \right) \ln\left(\frac{\sqrt{\rho}}{\eta b^2}\right) = \frac{1}{2(1-\nu)}\left(\BI - \nu \Be^{\beta}\otimes \Be^{\beta} \right) \ln\left(\frac{\rho}{\rho_0}\right),
\label{eq:ELineMS}
\end{equation}
where we used that the segment length (mesh length of the dislocation network) is now proportional to the characteristic dislocation spacing $(1/\sqrt{\rho})$. The parameter $\eta$ can again be adjusted to account for the dislocation core energy. There exists, in a three dimensional network, the possibility that a segment of a loop of slip system $\beta$ is collinear with a segment of slip system $\beta'$ (the two segments form a junction of Burgers vector $\Bb^{\beta\beta'}=\Bb^{\beta}+\Bb^{\beta'})$. We evaluate the junction energy as the sum of the energies of the constituent segments and an interaction energy. This interaction energy is strictly negative (otherwise the junction does not form). It is given by 
\begin{equation}
E_{J}^{\beta\beta'}  = \frac{\mu b^2}{4 \pi} g^{\beta\beta'}_{\rm J}\;,\quad
g^{\beta\beta'}_{\rm J} = \Bl^{\beta\beta'}.\left(\Bg_L^{\beta\beta'}-\Bg_L^{\beta}-\Bg_L^{\beta'}\right).\Bl^{\beta\beta'}.
\label{eq:ELineJ}
\end{equation}
where $\Bl^{\beta\beta'}$ is the direction of the junction segment which is constrained to form along the line of intersection of the slip planes of both slip systems. At first glance, our method of book-keeping may look unusual - why not directly evaluate the self energy of the junction segment and discarding the addition-cum-subtraction of the constituent segment energies? The reasons for this procedure will become transparent later.   

Discrete dislocation densities are now defined separately for each slip system as
\begin{equation}
\rho_{\rm d}^{\beta}(\Br,\phi^{\beta}) = \sum_{\cal S\in\beta} \int_{\cal S} 
\delta(\Br - \Br(s)) \delta(\phi^{\beta} - \phi^{\beta}(s)) \rmd s
\end{equation}
where $\phi^{\beta}$  is the angle between the Burgers vector $\Bb^{\beta}$ and the line direction $\Bl(s)$.  For junction segments we define junction densities 
\begin{eqnarray}
\rho_{\rm j}^{\beta\beta'}(\Br,\phi^{\beta}) &=& \sum_{\cal S\in(\beta,\beta')} \int_{\cal S} 
\delta(\Br - \Br(s)) \delta(\phi^{\beta} - \phi^{\beta\beta'}) \rmd s  = \rho_{\rm d}^{\beta}(\Br,\phi^{\beta}) f^{\beta\beta'}(\Br,\phi^{\beta})\nonumber\\
& = &
\nonumber\\
\rho_{\rm j}^{\beta'\beta}(\Br,\phi^{\beta'}) &=& \sum_{\cal S\in(\beta,\beta')} \int_{\cal S} 
\delta(\Br - \Br(s)) \delta(\phi^{\beta'} - \phi^{\beta'\beta}(s)) \rmd s  = \rho_{\rm d}^{\beta'}(\Br,\phi^{\beta'}) f^{\beta'\beta}(\Br,\phi^{\beta'})
\end{eqnarray}
Here, the function $f^{\beta\beta'}$ has the value $1$ whenever a segment of slip system $\beta$ forms a junction with a segment of slip system $\beta'$, and the value 0 otherwise. Note that a junction can alternatively be envisaged as a segment of orientation $\phi^{\beta\beta'} = \arccos(\Bl^{\beta\beta'}\Be^{\beta})$ in slip system $\beta$ or as a segment of orientation 
$\phi^{\beta'\beta} = \arccos(\Bl^{\beta\beta'}\Be^{\beta'})$ in slip system $\beta'$. 

For describing interactions between non-collinear dislocation segments, we define pair densities for pairs of slip systems as
\begin{equation}
\rho_{\rm d}^{\beta\beta'}(\Br,\phi^{\beta},\Br',\phi^{\beta'}) = \sum_{{\cal S \in \beta}\neq{\cal S'} \in\beta'} \iint_{\cal SS'} \delta(\Br - \Br(s)) \delta(\Br' - \Br(s')) \delta(\phi^{\beta} - \phi^{\beta}(s))\delta(\phi^{\beta'} - \phi^{\beta'}(s')) \rmd s \rmd s'.
\end{equation}
Upon averaging, all these densities become continuous functions of their arguments and we drop the subscript d. As previously we write the pair density functions in terms of products of single dislocation densities and pair correlation functions, 
\begin{equation}
\rho^{\beta\beta'}(\Br,\phi^{\beta},\Br',\phi^{\beta'}) = \rho^{\beta}(\Br,\phi^{\beta})\rho^{\beta}(\Br,\phi^{\beta})[1+
d^{\beta\beta'}(\Br,\phi^{\beta},\Br',\phi^{\beta'})]
\end{equation}
We skip the intermediate steps which proceed in direct analogy with those for a single slip system, with the only differences that now we need to sum over all slip systems (for the self energy) and all pairs of slip systems (for the interaction energy), and that we need to account explicitly for the junction energy. As in the previous section, we expand the slip system specific dislocation densities and correlation functions into alignment tensors
\begin{eqnarray}
\Brho^{\beta,[n]}(\Br) &=& \int \rho^{\beta}(\Br,\phi^{\beta}) \Bl(\phi^{\beta}) [\otimes \Bl(\phi^{\beta})]^{n-1} \rmd \phi{\beta}. \\
\Bd^{\beta\beta'[n,m]}(\Br,\Br') &=& \int d(\Br,\Br',\phi,\phi') \Bl(\phi^{\beta}) [\otimes \Bl(\phi^{\beta})]^{n-1} \otimes \Bl(\phi^{\beta'})[\otimes \Bl(\phi^{\beta'})]^{m-1} \rmd \phi^{\beta} \rmd \phi^{\beta'}. 
\label{eq:rhodalign}
\end{eqnarray}
Using these notations we write the self and Hartree energies as 
\begin{eqnarray}
E_{\rm S} &=& \frac{\mu b^2}{4 \pi} \sum_{\beta}\int \Brho^{\beta,[2]}:\Bg_L^{\beta} \rmd^3 r\\
E_{\rm H} &=& \frac{\mu b^2}{8 \pi} \sum_{\beta\beta'}\iint 
\Brho^{\beta,[1]}(\Br).\Bg^{\beta\beta'}(\Br - \Br').\Brho^{\beta',[1]}(\Br') \rmd^3 r\rmd^3 r'
\end{eqnarray}
where $\Bg^{\beta\beta'}$ is obtained from Eq. (\ref{eq:g}) by setting $\Be^{(i)} = \Be^{\beta},\Be^{(j)} = \Be^{\beta'}$. A new contribution to the system energy in case of multiple slip systems is the junction energy. To represent the junction energy we introduce the definition 
\begin{equation}
f^{\beta\beta'} = h^{\beta\beta'} \frac{\rho^{\beta'}}{\rho}
\end{equation}
with $h^{\beta\beta'}=h^{\beta'\beta} \le 1$ to meet the conditions $\sum_{\beta'} f^{\beta\beta'} < 1$ (only a fraction $< 1$ of all dislocation segments can form junctions) and $f^{\beta\beta'} \rho^{\beta} = f^{\beta'\beta} \rho^{\beta}$ (a junction between segments of slip systems $\beta$ and $\beta'$ is a junction of slip systems $\beta'$ and $\beta$). Using this notation we can write the junction energy as
\begin{equation}
E_{\rm J} = \frac{\mu b^2}{8 \pi} \sum_{\beta\beta'}\int \frac{\rho^{\beta}(\Br)\rho^{\beta'}(\Br)}{\rho(\Br)}
h^{\beta\beta'}(\Br)g^{\beta\beta'}_{\rm J} \rmd^3 r\\
\label{eq:EJ}
\end{equation}
In this expression we have made the simplifying assumption that the probability of forming a junction does not depend strongly on the orientation of the intersecting dislocations in their respective slip planes. A more general treatment which uses an alignment tensor expansion of $h^{\beta\beta'}$, and of which Eq. (\ref{eq:EJ}) is the lowest-order term,  will be given elsewhere. 

We are left with evaluating the correlation energy which contains all terms dependent on the correlation functions $d^{\beta\beta'}$. This can be written as
\begin{equation}
E_{\rm C} = \frac{\mu b^2}{8\pi} \sum_{\beta\beta'} \sum_{n,m=1}^{\infty} \iint \rho^{\beta,[n]}(\Br)  \stackrel{(n)}{:} \BT^{\beta\beta',[n,m]} \stackrel{(m)}{:} \rho^{\beta'[m]}(\Br) \rmd^3 r \\ 
\label{eq:ECLOCMS}
\end{equation}
where the interaction coefficients are 
\begin{equation}
\BT^{\beta\beta',[n,m]} = \frac{\BD^{\beta\beta'[n,m]}}{\rho} \quad,
D^{\beta\beta'[n,m]}_{i_1\dots i_n,j_1\dots j_m} = \frac{2^{m+n-2}}{4\pi^2} \int \tilde{d}^{\beta\beta'[n-1,m-1]}_{i_1\dots i_{n-1}, j_1 \dots j_{n-1}}(\Bu)
g^{\beta\beta'}_{i_n,j_m}(\Bu) \rmd^3 u. 
\label{eq:ECoeffDMS}
\end{equation}
with $\Bu = \Br \sqrt{\rho}$.
We note that the main qualitative difference between the single and multiple slip situations resides in the possible existence of collinear segments, i.e. junctions. In comparison with mutual interactions between distant segments of different loops, junctions may lead to a much more efficient energy reduction. 

\section{Applications}

\subsection{Dislocation screening in two dimensions}

As an application of our two-dimensional theory, we revisit the problem of Debye screening of dislocations which has been previously studied by Groma and co-workers \cite{Groma2006}. We use the energy functional given by Eq. (\ref{eq:Etot2Dfinal}) to evaluate the reponse of a homogeneous, infinitely extended 2D dislocation system of density $\rho$ to a single excess dislocation fixed in the origin, 
$\kappa_{0}(\Br) = \delta(\Br)$. The induced excess dislocation density $\kappa$ follows by considering the variation of the ensuing energy functional, under the assumption that the overall density $\rho$ remains homogeneous. The variation of Eq. (\ref{eq:Etot2Dfinal}) with respect to $\kappa$ is then given by
\begin{eqnarray}
\delta E = \frac{\mu b^2}{4\pi(1-\nu)}\int\left[D_{\rm II} \frac{\kappa}{2\rho} + \int [\kappa(\Br') + \delta(\Br')]g(\Br - \Br') \rmd^2 r'\right]\delta \kappa(\Br) \rmd^2 r \stackrel{!}{=}0.
\label{eq:2Dvar}
\end{eqnarray}
This leads to the following equilibrium equation for the induced density $\kappa$:
\begin{eqnarray}
\frac{\mu b^2}{4\pi(1-\nu)} \left[D_{\rm II} \frac{\kappa}{2\rho} + \int \kappa(\Br') g(\Br-\Br') \rmd^2 r' + g(\Br)\right] = 0.
\label{eq:2DvarII}
\end{eqnarray}

The solution of Eq. (\ref{eq:2DvarII}) can be found by Fourier transformation. Using
\begin{equation}
g(\Bk)= \frac{8\pi k_y^2}{k^4}
\end{equation}
and the definition $k_0^2 = 4 \pi\rho/D_{\rm II}$, we find 
\begin{equation}
\kappa(\Bk) = - \frac{4 k_0^2 k_y^2}{4k_0^{2}k_y^2 + k^4}
\label{eq:2DvarFT}
\end{equation}
from which reverse Fourier transformation yields the result 
\begin{equation}
\kappa(\Br) = \frac{k_0^2}{\pi} \left[\frac{y \sinh(k_0 y)}{r} K_1(k_0 r) -  \cosh(k_0 y) K_0(k_0 r)\right]\;.
\label{eq:2Ddebye}
\end{equation}
This is also the result obtained by Groma and co-workers \cite{Groma2006} and, using a quite different formalism, by Linkumnerd and Van der Giessen \cite{Limkumnerd2008}. We point out that our investigation, though it leads to the same result, differs somewhat from the work of Groma and also of Limkumnerd and Van der Giessen. Groma et. al use a free energy functional which is devised heuristically and the term $D$, which controls the range of correlations, is associated with entropy-like terms in the free energy. The same is true for the investigation of Limkumnerd and Van der Giessen \cite{Limkumnerd2008} who relate the range of correlations to fluctuation terms in the dislocation dynamics which they characterize by an effective temperature. In the present investigation, on the other hand, the parameter $k_0$, or $D_{\rm II}$, which controls the interaction range, arises from purely energetic considerations. Comparison with discrete simulations gives $k_0 = 4.2 \sqrt{\rho}$ \cite{Groma2006} which allows us to obtain the numerical value of the coupling parameter $D_{\rm II} \approx 0.84$. Inserting this numerical value into Eq. (\ref{eq:Etot2Dfinal}) together with typical values $\rho \approx 10^{12}$m$^{-2}$ and $\rho_0 \approx b^{-2} \approx 10^{19}$m$^-2$ indicates that, in the absence of long-range stresses (Hartree energy), the composition of the dislocation arrangement (geometrically necessary vs. statistically stored dislocations) has only a quite modest influence on the energetics. Even in the extreme limit $\kappa = \rho$ (only geometrically necessary dislocations) the additional energy cost implicit in the term proportional to $\kappa^2$ amounts only to about 3\% of the term proportional to $\rho$.

\subsection{Derivation of 'back stress' terms in two and three dimensions}

In dislocation-based plasticity theories, many authors have found it convenient to introduce 'back stress' terms proportional to the gradient of the dislocation density vector $\Bkappa$ into the stress balance, see e.g. \cite{Groma2003,Yefimov2004,Bayley2006,Roy2008,Luscher2015}. Such terms are of interest also because the dislocation density vector $\Bkappa$ is proportional to the gradient of plastic strain, hence, 'back stress' terms correspond to second-order plastic strain gradients entering the stress balance, a device highly popular in phenomenological gradient plasticity models of continuum mechanics. We demonstrate in this section that such terms arise naturally from our density functional representation of the dislocation energy. 

To this end, we first consider the 2D case. We take the $\kappa$-dependent terms in the energy functional given by Eq. (\ref{eq:Etot2Dfinal}) and insert the relation between the excess dislocation density $\kappa$ and the plastic strain $\gamma$, $\kappa = -(1/b) \partial_x \gamma$:
\begin{eqnarray}
E = \int \frac{\mu D_{\rm II}}{16\pi(1-\nu)\rho} (\partial_x \gamma)^2 \rmd^2 r + \frac{\mu}{8\pi(1-\nu)} \int
\partial_x \gamma(\Br) \partial_x \gamma(\Br')g(\Br - \Br') \rmd^2 r \rmd^2 r' \;.
\label{eq:Etot2Dgamma}
\end{eqnarray}
Variation with respect to $\gamma$ yields
\begin{equation}
\delta E = \int \frac{\mu}{4\pi(1-\nu)\rho} \left[\frac{D_{\rm II}}{2}\partial_x^2 \gamma  - \int
\partial_x \gamma(\Br') \partial_x g(\Br - \Br') \rmd^2 r'\right]\delta \gamma \rmd^2 r  \stackrel{!}{=} \int \tau(\Br) \delta \gamma(\Br) \rmd^2 r
\label{eq:Etot2Dvar}
\end{equation}
where we have used that the work conjugate of the plastic shear strain $\gamma$ is a resolved shear stress in the considered slip system. Using that the shear stress of a single dislocation is given by $\tau_{\rm d}(\Br) = - \mu b/(4 \pi(1-\nu)) \partial_x g(\Br)$ we see that the shear stress in the slip system is of the form
\begin{equation}
\tau(\Br) = - \frac{\mu b D_{\rm II}}{8\pi(1-\nu)\rho} \partial_x \kappa  + \int
\kappa(\Br') \tau_{\rm d}(\Br - \Br') \rmd^2 r' = \tau_{\rm b} + \tau_{\rm sc}
\end{equation}
The first of these terms is the back stress $\tau_{\rm b}$ derived, along a quite different line of reasoning, by Groma et. al \cite{Groma2003}. The present derivation makes it obvious that this term results from the correlation energy contribution that is quadratic in the excess dislocation density $\kappa$. The second term, $\tau_{\rm sc}$, represents the superposition of the long-range stress fields of the excess dislocations. This stress contribution derives from the Hartree energy and is normally obtained from solving the standard elastic-plastic problem. 

We can repeat the same argument for 3D systems. In case of a single slip system, the dislocation density vector relates to the strain gradient by $\Bkappa = (1/b) \Bepsilon_{\Bm n}\nabla \gamma$, $\Bepsilon_{\Bm n} = \Bepsilon.\Bn$ where $\Bepsilon$ is the Levi-Civita tensor and $\Bn$ the slip plane normal. The tensor $\Bepsilon_{\Bm n}$ rotates a vector contained in the slip plane, such as $\Bkappa$, counter-clockwise by $90^{\circ}$. With this notation we can write the $\Bkappa$-dependent terms in the energy functional, Eq. (\ref{eq:ECDD2}), as
\begin{eqnarray}
E(\Bkappa) 
&=& \frac{\mu b^2}{8\pi} \iint \Bkappa(\Br). \Bg(\Br-\Br').\Bkappa(\Br') \rmd^3 r \rmd^3 r'
+ \frac{\mu b^2}{8\pi} \int  \frac{\Bkappa(\Br). \BD^{[1,1]}.\Bkappa(\Br)}{\rho(\Br)} \rmd^3 r\nonumber\\
&=&
\frac{\mu b^2}{8\pi} \iint  [\Bepsilon_{\Bm n}.\nabla\gamma(\Br')].\Bg(\Br-\Br').[\Bepsilon_{\Bm n}.\nabla\gamma(\Br)] \rmd^3 r \rmd^3 r'
+ \frac{\mu b^2}{8\pi} \int  \frac{[\Bepsilon_{\Bm n}.\nabla\gamma(\Br).] 
\BD^{[1,1]}.[\Bepsilon_{\Bm n}.\nabla\gamma(\Br)]}{\rho(\Br)} \rmd^3 r
\end{eqnarray}
Variation with respect to $\delta \gamma$ gives
\begin{eqnarray}
\delta E(\Bkappa) &=& \frac{\mu b^2}{4\pi} \int \left[\int [\Bepsilon_{\Bm n} \nabla \gamma(\Br')] [\Bepsilon_{\Bm n} \nabla\Bg(\Br-\Br')]
\rmd^3 r'
- \frac{\mu b^2}{4\pi} \int  \frac{[\Bepsilon_{\Bm n}\nabla] \BD^{[1,1]}.[\Bepsilon_{\Bm n} \nabla \gamma]}{\rho(\Br)} \right] \delta \gamma(\Br) \rmd^3 r
\end{eqnarray}
The term in the brackets can again be understood as the resolved shear stress. In terms of $\Bkappa$ it is given by
\begin{eqnarray}
\tau(\Br) &=& \frac{\mu b}{4\pi} \int  \Bkappa(\Br') [\Bepsilon_{\Bm n} \nabla]_{\Br} \Bg(\Br-\Br') \rmd^3 r'
- \frac{\mu b}{4\pi\rho} [\Bepsilon_{\Bm n} \nabla].\BD^{[1,1]}.\Bkappa(\Br) =  \tau_{\rm b} + \tau_{\rm sc}
\end{eqnarray}
In case of multiple slip systems, we use the notation $\Bepsilon_{\Bm n}^{\beta} = \Bepsilon.\Bn^{\beta}$ where $\Bn^{\beta}$ is the slip plane normal of slip system $\beta$. After repeating the steps as above we get for the slip system specific back stress terms in multiple slip conditions
\begin{eqnarray}
\tau_{\rm b}^{\beta}(\Br) &=& - \frac{\mu b^2}{4\pi\rho} [\Bepsilon_{\Bm n}^{\beta} \nabla].[\sum_{\beta'} \BD^{\beta\beta'[1,1]}.\Bkappa^{\beta'}(\Br)].
\label{eq:taub3D}
\end{eqnarray}
Some implications of our derivation of the back stress term are discussed in Appendix C. 

\subsection{Estimate of the friction stress for a dislocation moving in a multiple slip environment}

Because of the geometrical constraints to dislocation glide on slip planes, dislocations can in general not move without intersecting dislocations on other slip systems. Because of this, sustained dislocation motion requires the repeated formation and breaking of junctions. The work required to break junctions is dissipated in the process. The distance between statistically equivalent configurations in the average direction of dislocation motion is given by the mesh length $1/\sqrt{\rho}$ of the dislocation network. Thus, the energy dissipated in advancing the dislocation by a distance $\delta u > 1/\sqrt{\rho}$ can be estimated as
\begin{equation}
E_{\rm diss}^{\beta}(\delta u) = \sum_{\beta'} \sum_{{\cal S}\in{\beta\beta'}} \int_{\cal S} E_J^{\beta\beta'} [\sqrt{\rho} \delta u(s)]  \rmd s =: \sum_{\cal S} \int_{\cal S} \tau^{\beta}(\Br) b \delta u(s) \rmd s 
\end{equation}
where $\tau^{\beta}$ is the friction stress required to move the dislocations, i.e., the resolved shear stress $\tau$ which provides the work required for breaking junctions. Upon transition to an averaged formulation we can re-write this equation as
\begin{equation}
E_{\rm diss}^{\beta} = \frac{\mu b^2}{4 \pi(1-\nu)} \sum_{\beta'}\int [\sqrt{\rho}(\Br) \frac{\rho^{\beta}(\Br)\rho^{\beta'}(\Br)}{\rho(\Br)}h^{\beta\beta'}g_J^{\beta\beta'}] \delta u(\Br) \rmd^3 r = \int \rho^{\beta}(\Br) \tau^{\beta}(\Br) \delta u(\Br) \rmd^3 r.
\end{equation}
Since the virtual displacement $\delta u$ is arbitrary, it follows that the shear stress ('friction stress') required to move the dislocation is
\begin{equation}
\tau^{\beta}_{\rm f}(\Br) = \frac{\mu b}{4 \pi(1-\nu)} \sum_{\beta'} \frac{\rho^{\beta'}(\Br)}{\rho(\Br)}
h^{\beta\beta'}(\Br)g^{\beta\beta'}_{\rm J} \sqrt{\rho(\Br)}.
\end{equation}
This stress obeys the generic Taylor scaling relation, i.e., it is proportional to the square root of dislocation density. It also depends on the distribution of dislocations over the various slip systems and on the coefficients $h^{\beta\beta'}g^{\beta\beta'}$ which have the character of latent hardening coefficients. The basic idea underlying the above argument is that the junction energy defines the amplitude of the small-scale energy fluctuations (on scales of the order of one dislocation spacing) that need to be overcome in order to move a dislocation by repeated breaking and formation of junctions. We note that the above argument can be generalized by replacing the scalars $h^{\beta\beta'}$ and $\rho^{\beta}$ with alignment tensor expansions of the corresponding angle-dependent functions. In this manner one can account for the fact that the average junction length may depend on the orientation distribution of the intersecting dislocations. This will be discussed in detail elsewhere. 

\section{Discussion and Conclusions}

It is interesting to compare our results with related work by other researchers, notably regarding the structure of the energy functional. We have shown that the energy functionals of dislocation systems possess a generic structure which is common to 2D and 3D dislocation systems. Specifically, the energy functionals consist of a 'Hartree' energy which is a non-local, quadratic functional of the dislocation density vector, or equivalently of the dislocation density tensor. This part of the energy functional does not depend on assumptions regarding dislocation correlations. The Hartree energy is complemented by an energy term which has the form $E_{\rm s} \propto - \int \mu b^2 \rho \ln (\rho/\rho_0) d^D r$ where $\rho_0 \propto (1/b^2)$. This energy is proportional to the line length per unit volume with a proportionality factor that decreases with increasing dislocation density, reflecting the fact that the screening radius of dislocation systems is proportional to the dislocation spacing. Terms of the form $ - \rho \ln \rho$ in a free energy density are normally associated with entropy, and indeed such terms appear in thermodynamic theories of dislocation systems, see e.g. Kooiman \cite{Kooiman2014}. However, in thermodynamic theories the pre-factor of $\rho \ln \rho$ type entropy terms is bound to be of the order of $kT$, which is several orders of magnitude less than the actual pre-factor $\approx \mu b^2$. The present derivation makes it clear that this term is in fact of energetic origin. We note that, in the hypothetical case where the dislocation density approaches $\rho_0$, according to the present formalism the energy per unit dislocation length in a system that is equally composed of positive and negative dislocations would go to zero. This is simply a reflection of the fact that in this case the cores of the positive and negative dislocations overlap and the dislocations annihilate. The role of the parameter $\rho_0$ is thus the exact opposite of the 'limit dislocation density' $\rho_s$ introduced in an ad-hoc manner by Berdichevsky \cite{Berdichevsky2006}. This term was introduced into the logarithmic factor in such a manner that it makes the energy per unit dislocation length  {\em diverge} as the dislocation density approaches the critical value $\rho_s$. In view of our results this idea must be discarded. Indeed, if we consider dislocation systems of zero net Burgers vector, it is difficult to see how densification of the dislocation system, i.e. bringing dislocations of positive and negative sign closer to each other, could conceivably increase rather than decrease the energy per dislocation length. The third energy contribution which consistently emerges from the present treatment is a local term which is quadratic in the excess (geometrically necessary) dislocation density. This term forms part of the 'correlation energy'; it depends on the structure of the dislocation pair correlation functions. Upon variation, it yields the 'back stress' which has become very popular in both phenomenological and dislocation density-based plasticity theories, not least because of its ability to explain size effects \cite{Groma2003,Yefimov2004}. To summarize, we hold the following fundamental structure of the energy density in the dislocation energy functional to be generic:
\begin{itemize}
\item Self energy terms of form $\propto - \mu b^2 \int \rho \ln (\rho/\rho_0) \rmd^D r$
\item A non-local Hartree energy which depends on the excess (GND) dislocation density vector, or equivalently on the dislocation density tensor, of form $\mu b^2 \iint \Bkappa(\Br). \Bg(\Br-\Br')  \Bkappa(\Br') \rmd^D r\rmd^D r'\;$ or $\mu b^2 \iint \Balpha(\Br). \BR(\Br-\Br') \Balpha(\Br') \rmd^D r \rmd^D r'$ 
\item Terms proportional to the square of the excess dislocation density vector, of form $\mu b^2\int {\ell}^2 \Bkappa.\BD.\Bkappa \rmd^D r$
\end{itemize}
Of these terms, the self energy depends only logarithmically (through the term $\rho_0$) on the correlation functions. Also dependent on the structure of the correlation functions are the length scale ${\ell}$ and the coupling tensors $\BD$. Of these we only know that they must be positively definite, since otherwise a homogeneous system of statistically stored dislocations would spontaneously decompose - which it does not. As to the length scale ${\ell}$, in the absence of other factors it must, in order to be consistent with the scaling properties of discrete dislocation systems \cite{Zaiser2001,Zaiser2014}, be chosen proportional to the dislocation spacing. This is the approach used in the present work. 

In our evaluation of the energy functional of a dislocation system we have made the key assumption that the range of correlations between dislocations is limited. This is  a necessary assumption for expressing the correlation energy as a {\em local} functional of the dislocation densities, see Appendix B. In physical terms this assumption corresponds to the simple idea that, by studying the dislocation configuration in one point, one cannot gain any information about the configuration of dislocations in a distant point (at a distance of many dislocation spacings) which is not contained in the slip-system dependent dislocation densities in that point. 
While this seems a rather mild assumption, it must be noted that a lot of dislocation systems that have been extensively investigated in the literature fall {\em not} into this category, among them:
\begin{itemize}
\item the Taylor lattice \cite{Hansen1986},
\item the infinite periodic or non-periodic dislocation wall \cite{Roy2008,Zaiser2013},
\item the infinite dislocation pile-up \cite{Schulz2014,Geers2013},
\item the periodic misfit dislocation array.
\end{itemize}
Many of these systems are one-dimensional and/or periodic, which makes powerful mathematical tools available for their analysis. The existence of these powerful tools may motivate the investigation but, on the other hand, it is also clear that most of the dislocation arrangements which develop during plastic deformation and whose properties govern plastic flow are disordered rather than ordered on large scales, and are two- or three-dimensional rather than one-dimensional. No Taylor lattice and no infinite periodic dislocation wall has ever been seen in the electron microscope, extended pile ups are the exception rather than the rule in deformation of real materials, and even in case of interface dislocations the assumption of periodic order has recently been called into question. We may thus argue that the study of low dimensional and/or periodic dislocation arrangements (which can never be captured by the present approach) is a consequence of mathematical convenience rather than of their practical importance. In this sense the present investigation can hopefully be considered a step in 'becoming generic' which also means 'becoming realistic'.

Among the applications we have given, we consider the derivation of back stress terms to be of fundamental interest. It should be clear from our derivation that the fundamental term in the energy functional that depends on the excess dislocation density $\Bkappa$ is the Hartree energy. Whatever assumptions are made regarding the correlation functions, this term is bound to stay. The term proportional to $\Bkappa^2$ in the correlation energy, which gives rise to back stress terms, is in fact a local correction to the fundamentally non-local functional -- a fact well recognized in recent work by Kooimans et. al. Accordingly, the back stress is a local correction to the in general non-local, long ranged interaction between excess dislocation densities in different parts of the crystal. In view of this fact it is astonishing that there are several published attempts to {\em replace}, rather than correct, the long-range dislocation interaction by a back stress term. This is tantamount to throwing out the Hartree energy and expressing the elastic energy as a functional of the excess dislocation density which does not contain any non-local, long-range interaction terms. The standard device used to this end is a truncation of the kernel $\Bg$ at some arbitrary radius ${\ell}$, see e.g. \cite{Bayley2006,Luscher2015}. We cannot help pointing out that this is inconsistent with the most fundamental property of dislocations, and of dislocation systems, namely the existence of a Burgers vector that is independent on the Burgers circuit: Let us compute the stress (or equivalently the elastic strain) associated with an arbitrary dislocation arrangement contained in some circle ${\cal C}_R$ of radius $R$. If we now truncate the stress around dislocations at length ${\ell}$, we are left with a horrible dilemma -- either we truncate the stress but not the strain, in which case we have destroyed elasticity, or we truncate both, in which case the integral over the circle ${\cal C}_{R+\ell}$ will yield zero whatever the Burgers vector content in ${\cal C}$  might be, which would be possible only if dislocations had no Burgers vector to begin with. This idea is taken to its logical conclusion in the work of Luscher et. al. \cite{Luscher2015} who use the back stress to evaluate the dislocation-associated strain as a compatible tensor field, see Appendix C. 

There are several directions how the present investigation could be expanded and further developed. At present, our treatment of segment self-interactions via a line energy approximation is not very elegant. This can be easily improved upon by replacing the interaction tensors $\Bg$ by core-regularized expressions which can be derived in various manners including gradient elasticty \cite{Lazar2005} and continuous Burgers vector distributions around the dislocation core \cite{Cai2006}. Since we cannot, in the general case, calculate dislocation correlation functions from our theory, we need to obtain the information regarding quantities like $\rho_0$ and $\BD^{[n,m]}$ from external sources. In our opinion, as a next step, a systematic effort is needed to evaluate the expansion parameters of the theory (the coupling tensors $\BD^{[n,m]}$) from DDD simulations. From a numerical point of view this is not difficult, especially with reference to DDD codes that express the interaction energy in terms of line integrals over the dislocation lines \cite{Ghoniem1999,Ghoniem2000}, since the coupling tensors can for discrete dislocation systems be evaluated in a similar manner. A comparison with DDD simulation data will serve two important objectives. Firstly, owing to the non-ergodicity of dislocation dynamics, it is not at all clear to which extent these coupling tensors depend on initial conditions. If they do so in a sensitive manner, meaning that different types of initial conditions lead to quite different values for the coupling tensors and thus to different energy functionals, then the present theory is useless for practical application. If, on the other hand, the dependence on initial conditions and deformation geometry is weak and remains within the statistical scatter among individual simulations, then the theory can be applied for evaluating the dynamics of dislocation systems from density-based evolution equations in a correspondingly wide range of situations. Secondly, comparison with DDD simulation can tell us how many terms of the alignment tensor expansion are actually needed for a meaningful representation of dislocation energetics, and thus provide important hints regarding the question what degree of complexity is actually needed for density based dislocation dynamics models. DDD simulations can be usefully complemented by experimental data regarding the structure of energy functionals for dislocation systems. Classical X-ray and calorimetry studies (see e.g. \cite{Ungar1984}) provide information about the energy stored in a dislocated crystal and, via line profile analysis, about the range and dislocation density dependence of screening correlations in dislocation systems (Wilkens' $M$-Parameter, \cite{Ungar1984}). Recent developments in X-ray microscopy allow to map the lattice distortions, and hence the elastic energy density associated with dislocation systems on scales well below the spacing of individual dislocations, see the impressive work of Wilkinson and co-workers \cite{Wilkinson2014}. Such experiments allow to obtain data that are of comparable quality to those from DDD simulation and can be used in a similar manner for evaluating the elastic energy functional and parameterizing its local and non-local terms.   

\acknowledgments
{We gratefully acknowledge financial support by for the research group FOR1650 Dislocation based plasticity funded by the German Research Foundation (DFG) under Contract Number ZA171/7-1.

\section*{Appendix}

\subsection{Averaging procedures}

To represent a discrete dislocation system by continuous densities, averaging procedures are required. Spatial averaging is commonly used in mechanics, but has the disadvantage that it does not preserve information about the relative positions of dislocations (or particles) with respect to each other, information which is essential for capturing the energetics. Hence, spatial averages are not normally used in statistical mechanics where one aims at deriving the average properties of systems from the dynamics and interactions of their discrete elements. The only averages normally used in statistical mechanics are (i) temporal averages along the trajectory of a system or (ii) instantaneous averages over ensembles of many systems. In thermal equilibrium, both types of averages are assumed to coincide (ergodicity). 
However, dislocation systems during plastic deformation are not in thermal equilibrium, and moreover their dynamics is strongly influenced by constraints (glide on crystallographic planes) which normally prevent them from fully exploring the phase space of possible configurations. Hence, dislocation motion tends to be non-ergodic. This leaves us with ensemble averaging as the only feasible averaging approach.

Initial conditions for a 2D dislocation dynamics simulation are provided by assigning initial positions and signs to $N$ dislocations. Any statistical rule for doing so explicitly or implicitly defines an initial N-particle density probability density function
\begin{equation}
p_N(\Br_1\dots \Br_N, s_1 \dots s_N, s_1 \dots s_N) \rmd^D r_1\dots \rmd^D r_N 
\end{equation}
which is the joint probability to find the first dislocation of sign $1$ at $\Br_1$, dislocation 2 of sign $2$ at $\Br_2$, etc. Obviously, the N-particle density function fulfils the normalization condition
\begin{equation}
\sum_{s_1 \dots s_N}  \int p(\Br_1\dots \Br_N, s_1 \dots s_N) \rmd^3 r_1\dots \rmd^D r_N  =  1\:.
\label{norm}
\end{equation}
An non-equilibrium ensemble is defined by its initial probability density function and evolution equations, hence, the rules for constructing initial conditions in a set of multiple DDD simulations can be considered to define an ensemble. This is true for both 2D and 3D simulations: Initial conditions for a 3D dislocation dynamics simulation can be understood as statistical rules for assigning initial positions and directions to $N$ segments in terms of a density function $p(\Br_1\dots \Br_N, \phi_1 \dots \phi_N)$, where those rules need to respect line connectivity and one makes the transition $N \to \infty$ as the segments are made to be arbitrarily short. 

Probabilities of lower order can be obtained by integrating over some of the coordinates. Of particular importance in the present study are the single-particle and pair probabilities defined by
\begin{eqnarray}
p_1(\Br,s) &=&  \sum_{s_2 \dots s_N} \int p_N(\Br, \Br_2 \dots \Br_N, s, s_2 \dots s_N) \rmd^D r_2\dots \rmd^D r_N,\nonumber\\
p_2(\Br,s,\Br',s') &=& \sum_{s_3 \dots s_N}  \int p_N(\Br,\Br',\Br_3 \dots \Br_N, s,s',s_3 \dots s_N) \rmd^D r_3\dots \rmd^D r_N.
\end{eqnarray}
These give, respectively, the joint probability of finding a dislocation  at $\Br$ and with sign $s$, and the joint probability of finding a dislocation pair at $(\Br,\Br')$ with signs $(s,s')$, irrespective of the positions and signs of all other dislocations. The probability for a dislocation at any position to have sign $s$
is given by $p(s) = N_s/N$ where $N_s$ is the number of dislocations of sign $s$. We write
\begin{equation}
p_1(\Br,s) = p(s) f_1^s(\Br)\quad,\quad 
p_2(\Br,\Br',s,s') = p(s)p(s') f_2^{ss'}(\Br,\Br')
\end{equation}
where $f_1^s(\Br)$ is the conditional probability density for a dislocation of sign $s$ to be at $\Br$ and $f_2^{ss'}(\Br,\Br')$ is the conditional probability density for a dislocation pair of signs $(s,s')$ to be at the positions $(\Br,\Br')$. From the general normalization condition, Eq. (\ref{norm}), we see that these conditional probabilities are normalized according to $\int f_1^{s}(\Br) \rmd^D \Br = 1$, $\iint f_2^{ss'}(\Br,\Br') \rmd^D r\rmd^D r' = 1$. 

From the sign-conditional single-dislocation and pair probability densities we obtain the respective dislocation densities by
\begin{eqnarray}
\rho_s(\Br) &=&  N_s f_1^{(s)}(\Br) \quad,\quad
\rho_{ss'}(\Br,\Br') = \left\{ 
\begin{array}{l}
N_s N_{s'} \;f_2^{ss'}(\Br,\Br') \quad,\quad s' \neq s\\
N_s (N_s-1)\;f_2^{ss}(\Br,\Br') \quad,\quad  s' = s. 
\end{array} 
\right.
\end{eqnarray}
Note that, while  the number of pairs of dislocations of types $s \neq s'$ is $N_s N_{s'}$,  the number of pairs of dislocations of type $s$ is $N_s(N_s -1)$ since a dislocation cannot form a pair with itself. The densities are thus normalized to yield, upon spatial integration, the respective numbers of dislocations or dislocation pairs. From this normalization it follows that, if we express the pair density as $\rho_{ss'}(\Br,\Br') = \rho_{s}(\Br) \rho_{s'}(\Br') [1+d_{ss'}(\Br,\Br')]$ then
\begin{equation}
\int \rho_s(\Br) d_{ss'}(\Br,\Br') \rmd^2 r = \int \rho_{s'}(\Br') d_{ss'}(\Br,\Br') \rmd^2 r' = \left\{
\begin{array}{l}
0\quad,\quad s \neq s'\\
- 1 \quad,\quad s = s'.
\end{array} 
\right.
\end{equation}
We see that the quantity $\rho_s(\Br) d_{ss}(\Br,\Br')$ has a role similar to the exchange-correlation hole density in density functional theories of electron systems
(see e.g. \cite{Evarestov2007}). In a local density approximation where $\rho_s(\Br)$ depends only weakly on $\Br$ over the range of the correlation function $d_{ss}(\Br,\Br')$, we may pull the factors $\rho_s(\Br) \approx \rho_s(\Br') $ out of the integrals, and Eq. (\ref{eq:dppm}) follows. 

In conclusion, a point about the choice of initial conditions (the intial many-dislocation density function, or the initial conditions in a series of DDD simulations) is appropriate. To enable comparison with experiment, initial conditions in DDD should be consistent with information about dislocation microstructure that is accessible by experiment. In well characterized microstructures, this is typically the total dislocation density, the geometrically necessary density as monitored by lattice rotations or misorientations, and possibly the distribution of dislocations over the various Burgers vectors. In extremely well characterized specimens, even information about the distribution of dislocations over edge and screw orienations may be available.

All these informations are comprised in the dislocation density alignment tensors up to order two. However, any DDD simulation involves constructing initial conditions which imply the definition of a many-dislocation or many-segment density function - a function which may contain much more information than is contained in the simple density functions. It is in the opinion of the present author essential that initial conditions are constructed in a manner that does not introduce more information than is actually available, since otherwise the results may be influenced in a significant, and potentially uncontrollable, manner by hidden parameters introduced in the form of assumptions about the initial state that are not backed up by experimental evidence. A systematic manner of constructing unbiased initial conditions is provided by the maximum entropy method, which allows to construct the many-particle density by maximizing the entropy while using the available information as constraints. As an example, for a 2D dislocation system of size $L^2$ with $N+$ positive and $N-$ negative dislocations (densities $\rho_{\pm} + N^{\pm}/L^2$), the N-particle probability density function which maximizes the entropy is simply $p_N = (1/L^2)^{N+}(1/L^2)^{N-}$ -- in simple words, as an initial condition, the dislocations are placed independently at random locations, which is indeed a popular initial state for 2D DDD simulations. Other micro-arrangements which are popular in the literature for analysing properties of dislocation systems, for instance placing dislocations on regularly spaced slip planes \cite{Roy2008}, or even on a regular Taylor lattic \cite{Hansen1986}, seem highly problematic from an information-theoretical of view because they imply strong assumptions about a correlation structure which may not be backed up by experimental evidence. 

\subsection{Non-local density functional approximations of the correlation energy}

The local density approximation used in this work can be considered the lowest order of a systematic expansion of the energy functional in terms of gradients of the dislocation densities. We illustrate this for the correlation energy of a 2D dislocation system. We start from Eq. (\ref{eq:Etot2D}):
\begin{equation}
E_{\rm C}=\frac{1}{2} \sum_{ss'} \iint \rho_s(\Br)\rho_{s'}(\Br')d_{ss'}(\Br,\Br')E_{\rm int}(\Br - \Br')  \rmd^2 r \rmd^2 r'\;.
\label{eq:EC2D}
\end{equation}
We introduce the vectors $\Br^{*} = (\Br + \Br')/2$ and $\Ba = \Br - \Br'$ and expand both $\rho$ and $\rho_{s'}$ 
around the point $\Br^*$:
\begin{equation}
\rho_s(\Br) = \sum_{n=0}^{\infty} \frac{1}{n!} \left[\frac{\Ba.\nabla_{\Br}}{2}\right]^n  \rho_s(\Br)|_{\Br^*}\quad,\quad
\rho_{s'}(\Br) = \sum_{m=0}^{\infty} \frac{1}{m!} \left[\frac{\Ba.\nabla_{\Br}}{2}\right]^m  \rho_{s'}(\Br)|_{\Br^*}.
\end{equation}
Inserting into the correlation energy gives
\begin{equation}
E_{\rm C}\frac{1}{2} \sum_{ss'} \sum_{n,m} \frac{1}{2^{n+m}n!m!} 
\iint d_{ss'}(\Br^*,\Ba)
E_{\rm int}(\Ba) \left[\frac{\Ba.\nabla_{\Br}}{2}\right]^n \rho_s(\Br^*) \left[\frac{\Ba.\nabla_{\Br}}{2}\right]^m\rho_{s'}(\Br^*) \rmd^2 a \rmd^2 r^*\;.
\label{eq:EC2DTaylor}
\end{equation}
We then introduce the gradient coefficient tensors $\BT_{ss'}^{(n+m)}$ with components 
\begin{equation}
T^{n+m}
_{ss', i_1\dots i_{n+m}}(\Br^*) = \frac{1}{2^{n+m}n!m!} \int a_{i_1}\dots a_{i_{n+m}} d_{ss'}(\Br^*,\Ba) E_{\rm int}(\Ba) d^3 a
\label{eq:gradcoeff}
\end{equation}
to write the correlation energy as
\begin{equation}
E_{\rm C}=\frac{1}{2} \sum_{ss'} \sum_{n,m} 
\int \nabla^n \rho_s(\Br) \stackrel{n}{:}\BT_{ss'}^{(n+m)}(\Br) \stackrel{m}{:} \nabla^m \rho_{s'}(\Br) \rmd^2 r\;.
\label{eq:EC2DGrad}
\end{equation}
Here, the tensorial $m$-th order dislocation density gradient $\nabla^m \rho_s$ is the rank-m-Tensor with components $\partial_{i_1}\dots\partial_{i_m} \rho_s$. Thus the correlation energy can be represented in terms of a gradient expansion of the dislocation densities, provided that the dislocation density functions can be differentiated to arbitrary order and that the gradient coefficient tensors of arbitrary order exist. A necessary and sufficient condition for this is a faster than algebraic decay of the correlation functions $d_{ss'}(\Br,\Br')$, which corresponds to the assumption of a macro-disordered dislocation arrangement.  

It is straightforward to generalize the above argument to 3D dislocation systems, however, the notation associated with a double expansion in real space and in angular coordinates is cumbersome so we refrain from giving explicit expressions. The local density approximation used in the remainder of this paper is just the lowest-order term of the above mentioned gradient expansion. The above argument demonstrates that this approximation can be systematically generalized to derive gradient-dependent expressions for the correlation energy of any desired order. 

\subsection{How not to understand back stresses}

Eq. \ref{eq:taub3D} relates the back stress on a slip system to directional derivatives of the dislocation density vector:
\begin{eqnarray}
\tau_{\rm b}^{\beta}(\Br) &=& - \frac{\mu b^2}{4\pi\rho} [\Bepsilon_{\Bm n}^{\beta} \nabla].[\sum_{\beta'} \BD^{\beta\beta'[1,1]}.\Bkappa^{\beta'}(\Br)].
\end{eqnarray}
Our derivation tells us that this stress enters into the stress balance alongside the standard non-local stress ('Hartree' stress). However, several authors \cite{Bayley2006,Roy2008,Luscher2015} have suggested to use the back stress in order to {\em replace} long-range dislocation interactions. To illustrate the implications, we follow Luscher et. al. \cite{Luscher2015} who use an expression exactly analogous to the above equation, though with a scalar coupling constant $D$. If this expression is assumed to fully describe the dislocation associated stress field, it is only natural to associate the back stress with a matching stress tensor, $\tau_{\rm b}^{\beta} = \BMM^{\beta} \Bsigma_{\rm b}$, and this with a strain through  
\begin{equation}
\Bepsilon_{\rm b} = \BC^{-1}:\Bsigma_{\rm b} = {\BC}^{-1}:(\BMM^{\beta})^{-1} \tau_{\rm b}
\end{equation}
where $\BC$ is Hooke's tensor. Luscher et. al. use this expression to evaluate the dislocation associated strain which is only logical, since there is no other stress associated with dislocations in their theory. To show the implications we look at a special case -- our 2D dislocation system with Burgers vector $\Bb = b \Be_x$ and slip plane normal $\Bn = \Be_y$ containing straight parallel dislocations of line direction $\Bl = \Be_z$. The dislocation density vector is $\Bkappa = \kappa \Be_z$, and the back stress $\tau_{\rm b} = - (\mu b^2 D)/(4\pi\rho) \partial_x \kappa$. We now consider a particular dislocation distribution - a blob of positive dislocations with density $\kappa(\Br) = \kappa_0 \exp(-r^2/\ell^2)$ and ask what is the Burgers vector contained in a circle ${\cal C}_R$ of radius $R$ and area ${\cal A}_R$ around the origin. According to the classical definition the curl of the plastic distortion, the Burgers vector associated with $\kappa$ follows from Stokes theorem as $\Bb = \Be_x \int_{{\cal A}_R} \kappa d^2 r$ which goes to a finite value as $R \to \infty$. If we instead evaluate the associated Burgers vector from the 'dislocation strain' $\Bepsilon_{\rm b}$, we get 
\begin{equation}
b_i = - (\mu b^2 D)/(4\pi\rho) C^{-1}_{ijkl}(M^{\beta})^{-1}_{kl} \oint_{{\cal C}_R} \partial_x \kappa  \rmd s_j.
\end{equation}
It is easily seen that the integral, rather than converging to a constant value, becomes exponentially small once the radius of the circle becomes much larger than $\ell$. It follows that, in such a theory, depending on the choice of the Burgers circuit an accumulation of excess dislocations may have no Burgers vector at all. We conclude that the back stress should better not be used to replace the long range dislocation interaction.  
\end{document}